\documentclass[aps,prc,showpacs,twocolumn,superscriptaddress,nofootinbib,preprintnumbers,floatfix]{revtex4-1}

\usepackage{longtable}
\usepackage{graphicx}
\usepackage{slashed}
\usepackage{dcolumn}
\usepackage{bm}
\usepackage{amsmath,amssymb}
\usepackage{epstopdf} 
\usepackage{subfigure}
\usepackage{tensor}
\usepackage{gensymb} 
\usepackage{cancel}
\usepackage{float}

\newcommand{\bea}{\begin{eqnarray}}
\newcommand{\eea}{\end{eqnarray}}
\newcommand{\beq}{\begin{equation}}
\newcommand{\eeq}{\end{equation}}

\usepackage[normalem]{ulem}  
\usepackage[dvips]{color} 

\renewcommand\sout{\bgroup \color{red} \ULdepth=-.5ex \ULset}

\begin{document}



\title{Lepton-Proton Two-Photon Exchange in  Chiral Perturbation Theory}


\author{Pulak Talukdar}
     \email[]{t.pulak@iitg.ac.in}
     \affiliation{Department of Physics, Indian Institute of Technology Guwahati, 
                 Guwahati - 781039, Assam, India.}%
\author{Vanamali C. Shastry}
     \email[]{vanamalishastry@gmail.com}
     \affiliation{Department of Education in Science and Mathematics, 
                 Regional Institute of Education Mysuru, Mysore - 570006, India.}%
\author{Udit Raha}
     \email[]{udit.raha@iitg.ac.in}
     \affiliation{Department of Physics, Indian Institute of Technology Guwahati, 
                 Guwahati - 781039, Assam, India.}%
\author{Fred Myhrer}
     \email[]{myhrer@mailbox.sc.edu}
     \affiliation{Department of Physics and Astronomy, 
                 University of South Carolina, Columbia, SC 29208, USA.}%





\begin{abstract}
\noindent 
We use heavy baryon chiral perturbation theory to evaluate the  two-photon exchange 
corrections to the low-energy elastic lepton-proton scattering at next-to-leading order 
accuracy, i.e., ${\mathcal O}(\alpha, M^{-1})$, including a non-zero lepton mass. We 
consider the elastic proton intermediate state in the two-photon exchange invoking
soft photon approximation. The infrared singular contributions are projected out using 
dimensional regularization. The resulting infrared singularity-free two-photon exchange 
contribution is in good numerical agreement with existing predictions based on standard 
diagrammatic soft photon approximation evaluations. 
\end{abstract}


\maketitle

\section{Introduction}
\label{intro}
During the past few decades electron-proton ($e$p) scattering experiments at various experimental 
facilities, e.g.,  BINP Novosibirsk, SLAC, DESY, Fermilab, CERN, JLab, MAMI, have provided great 
insights into the electromagnetic structure of the proton. 
The point-like nature of the electrons as well as the small value of the electromagnetic coupling 
make them ideal probes for investigating the internal structure of the proton. Polarized and 
unpolarized cross section measurements with ultra-relativistic electrons have yielded information 
on physical quantities, like electromagnetic form factors, parton distribution functions, and 
polarization asymmetries. However, in recent years the values of some low-energy physical quantities 
extracted from such experiments show discrepancies which are currently difficult to reconcile. The 
most contentious being the measurements of the {\it root-mean-square} (rms) proton radius  extracted 
from $e$p scattering data versus the ones from the CREMA collaboration measurements, which are      
high-precision muonic hydrogen Lamb-shift determinations, leading to about $5\sigma$ discrepancy with 
the previous accepted proton rms 
value~\cite{Pohl:2010zza,Antognini,Codata,Pohl:2013yb,Bernauer:2010wm,Zhan:2011ji,Mihovilovic:2016rkr}. 
We note that a very recent hydrogen Lamb-shift measurement~\cite{Bezginov:2019}, however, reported a 
result consistent with the CREMA measurement~\cite{Pohl:2010zza,Antognini}.
This so-called  ``proton radius puzzle'' together with the well-known discrepancy of the electric to 
magnetic form factor ratio ($G_E/G_M$) of the 
proton~\cite{Walker:1994,Bosted:1994tm,Milbrath:1998,Milbrath:1999,Jones:1999rz,Gayou:2001qt,Gayou:2002,Brash:2002,Arrington:2003df},
has resulted in a renewed vigor in the study of the structure of the proton both experimentally and 
theoretically (see e.g., Refs.~\cite{Pohl:2013yb,Perdrisat:2006hj,Arrington:2011dn,Carlson:2015jba} for 
recent reviews).   

The  proton  rms radius is determined from the proton's  electric form factor ($G_E$) which may be obtained 
from the measurement of the unpolarized elastic lepton-proton ($\ell$p) scattering cross section. The rms 
electric charge radius ($\sqrt{\langle r^2_E\rangle}$) is thereby extracted using the relation 
$\langle r_E^2\rangle = 6\frac{\partial G_E(Q^2)}{\partial Q^2}|_{Q^2=0}$ where, $Q^2$ is the four-momentum 
transfer. One of the challenges associated with the measurements of the $\ell$p cross section at low-energy 
or low-$Q^2$ values, is the bremsstrahlung  process, $\ell$p$ \to \ell$p$\gamma$, which constitutes an 
important background. The data analysis entails the disentanglement of this background from the elastic 
$\ell$p scattering process before the rms radius can be extracted. To yield meaningful results one needs to 
deal with the {\it soft} photon emissions,  leading to infrared (IR) divergences that must cancel with the 
IR-divergent virtual photon exchange counterparts. This so-called {\it unfolding procedure} of the radiative 
analysis of the {\it raw} data~\cite{Tsai:1961zz,Mo:1968cg,Maximon:2000hm,Vanderhaeghen2000,Gramolin2014}, 
makes an experimental determination of the rms radius rather intricate especially at low-energies.  
	
Several recent experimental proposals, including low-energy $\ell$p scattering experiments, are under way 
to resolve the rms discrepancy. For example, the Prad~\cite{Prad,Prad2} experiment at JLab and the  MUon 
proton Scattering Experiment (MUSE)~\cite{Gilman:2013eiv,Gilman:2013vma}  at PSI are two such experiments. 
In particular, MUSE collaboration aims to measure the elastic $\mu^\pm$p scattering cross sections at momentum 
transfer as low as $|Q^2| \sim 0.002 - 0.08 \text{ GeV/c}^2$~\cite{Gilman:2013eiv,Gilman:2013vma}. In fact, 
MUSE plans to extract the proton's rms radius from very precise measurements (with a projected accuracy of 
less than $1\%$) of the $\mu^\pm$p and $e^\pm$p cross sections. This should facilitate a comparative study 
of the extracted rms radii from these low-energy elastic scattering processes under very similar experimental 
conditions. 
	
The proton is an extended particle composed of quarks and gluons, and for low-energy probes 
one is faced with complexities arising from the underlying non-perturbative nature of strong interactions. This 
low-energy theory is usually parametrized; the proton-photon vertices are described  by electric and magnetic 
(Sachs) form factors. The form factors are either phenomenologically modeled, extracted directly from experimental 
data, or determined via \textit{ab initio} numerical calculations using Lattice QCD. Well-known works from the 
past~\cite{Tsai:1961zz,Mo:1968cg,Maximon:2000hm}, as well as many recent works in the last two decades (e.g., 
\cite{Perdrisat:2006hj,Arrington:2011dn,tomalak2014two,Carlson:2015jba,Afanasev:2017gsk} and other 
references therein) on radiative correction analyses, relied on such phenomenological form factors. In contrast, 
our analysis presented in this work makes use of elementary point-like vertices derived in the context of an 
{\it effective field theory} (EFT). The work of Tsai~\cite{Tsai:1961zz} presented a detailed account of the radiative 
correction analysis for the elastic electron-proton scattering process where the relativistic recoil corrections for 
the proton were considered. This analysis therefore predominantly concentrated on the high-energy regime of the lepton 
scattering process. In a later work, Mo and Tsai~\cite{Mo:1968cg} introduced the so-called {\it peaking approximation} 
which is justifiable for electron scattering off the proton even at low energies, as confirmed in, e.g., 
Ref.~\cite{Talukdar:2018hia,Myhrer:2018ski}. 

One of the earliest works on the {\it two-photon exchange} (TPE) effects may be the so-called 
{\it Feshbach corrections}~\cite{McKinley:1948zz}, which considered relativistic electrons scattering off a 
static Coulomb potential. The later works of Refs.~\cite{Tsai:1961zz,Mo:1968cg,Maximon:2000hm} did consider the 
virtual TPE diagrams in order to cancel the IR divergences arising from the bremsstrahlung diagrams. These 
calculations suggested that the TPE effects were small. In other words, 
the dominant contributions arose from the 
{\it one-photon exchange} contribution (i.e., the first Born approximation) leading to the celebrated Rosenbluth 
formula for elastic lepton-proton scattering cross section. Modern experimental arrangements like the MUSE facilitate 
simultaneous measurement of the unpolarized elastic $e^\pm$p and $\mu^\pm$p scattering cross sections, thereby 
enabling extraction of possible enhanced TPE contributions. In other words, MUSE will measure the difference of the 
lepton and anti-lepton charge cross sections to which the interference between the Born and the TPE diagrams at 
$\mathcal{O}(\alpha^3)$ contributes.\!\!~\footnote{These {\it charge-asymmetry} measurements can not be used to extract 
the TPE contribution directly. The MUSE experiment can instead observe the {\it charge odd} combinations of  
TPE along with parts of the bremsstrahlung contributions. In order to isolate the TPE contribution, model-dependent 
corrections must be applied to the charge-asymmetry data, i.e., one has to  extract the charge-dependent bremsstrahlung 
contributions, e.g., Ref.~ \cite{Arrington:2011dn}. }

Recent theoretical studies have suggested that the TPE effects can play crucial role in explaining possible 
discrepancies in various measured observables. It appears to be the general consensus that the TPE contributions 
have the correct sign and magnitude in order to resolve the bulk of the discrepancies in the extraction of form 
factors~\cite{Walker:1994,Bosted:1994tm,Milbrath:1998,Milbrath:1999,Jones:1999rz,Gayou:2001qt,Gayou:2002,Brash:2002,Arrington:2003df,Pohl:2013yb,Perdrisat:2006hj,Arrington:2011dn}. This brought about a renaissance in TPE studies relating to the 
$\ell$p scattering process. A wide variety of hadronic model analyses of the TPE contributions include dispersion 
theory methods~\cite{Carlson:2007sp,Borisyuk:2006fh,Borisyuk:2008es,Borisyuk:2010cv,Gorchtein:2006mq,Belushkin:2006qa,Belushkin:2007zv,Lorenz:2012tm,Hoferichter:2016duk,Kuraev:2008gw,tomalak2015three,Tomalak:2015aoa,Tomalak:2015hva,Tomalak:2017npu,tomalak2018}~\!\!\footnote{{\it En passant},  Refs.~\cite{Belushkin:2006qa,Belushkin:2007zv} used dispersion relations to predict 
a smaller value of the proton's charge radius $\sim 0.84-0.85$ fm, prior to the CREMA muonic hydrogen 
measurements~\cite{Pohl:2010zza,Antognini}. }, resonance exchange models and dynamical coupled channel K-matrix 
analyses~\cite{Kondratyuk:2001qu,Penner:2002md,Korchin:2003ah,Kondratyuk:2005kk,Kondratyuk:2007hc,Zhou:2009nf}. 
In these evaluations of the TPE processes, inelastic intermediate states 
of the nucleon and the $\Delta$, namely, the $N^*$, $\Delta^*$ and other possible excitation, along with various 
resonance exchanges, such as  $\sigma,\, \omega,\, \phi$, were considered which could contribute even at 
small momentum transfers, $|Q^2|\lesssim 0.1$ (GeV/c)$^2$~\cite{Arrington:2011dn}. Contributions from these 
intermediate excited states are expected to be small at such low $|Q^2|$. Moreover, there has been a report of an 
interesting interplay between the spin-1/2 and spin-3/2 resonance states leading to partial cancellations among the 
above excited states of nucleon and $\Delta$ contributions to the TPE~\cite{Kondratyuk:2005kk}. Ultimately, the TPE 
with the elastic proton intermediate state is expected to give the dominant contribution at very low momentum 
transfers~\cite {tomalak2014two,Tomalak:2015hva,tomalak2015three,Tomalak:2015aoa,Tomalak:2017npu,BlundenPRL,Koshchii:2017dzr}. 
In this work we focus only on the intermediate elastic proton contributions to the TPE diagrams. Furthermore, here we 
only need to deal with the real parts of these amplitudes which contribute to the unpolarized elastic lepton-proton 
cross section. 

As already mentioned, the TPE contributions contain IR divergences which are canceled by the IR terms arising 
from the soft photon bremsstrahlung process at $\mathcal{O}(\alpha^3)$. In this work, we present an evaluation of 
TPE contributions with a proton intermediate state using a low-energy EFT, namely the Heavy Baryon Chiral 
Perturbation Theory (HB$\chi$PT), which is an effective low-energy field theory of QCD (e.g., 
\cite{Bernard:1995dp,Bernard:2007} and references therein). The primary motivation for the use of HB$\chi$PT 
is to provide a systematic, model independent evaluation of the TPE intermediate proton contribution at low energies 
incorporating simultaneous radiative and proton recoil effects. HB$\chi$PT entails a perturbative expansion of the 
chiral Lagrangian based on a momentum expansion scheme. The {\it leading chiral order} (LO) terms give the dominant 
amplitudes, and the {\it next-to-leading order} (NLO)  amplitudes normally, viz., in a naive dimensional analysis, 
are smaller corrections to the LO amplitudes of the process. HB$\chi$PT also includes a well established perturbative 
counting  expansion in inverse powers of the nucleon mass $M$ consistent with the chiral momentum expansion. Since 
the chiral symmetry {\it breakdown scale} $\Lambda_\chi$ is of the order of $M\sim 1$ GeV, the expansion parameter, 
$Q/\Lambda_\chi\sim m_\pi/\Lambda_\chi \ll 1$, includes both the chiral expansion and the expansion in $M^{-1}$. Moreover, 
the electromagnetic interaction naturally enters HB$\chi$PT in a gauge invariant way. 

Thus, HB$\chi$PT provides the ideal framework to study low-energy processes like the $\ell$p scattering, 
where nucleons, mesons and leptons are the fundamental degrees of freedom. Especially in dealing with 
MUSE-like kinematics where the lepton mass plays a sensitive role, the widely used ultra-relativistic 
approximation of leptons can not be employed~\cite{Talukdar:2018hia,Myhrer:2018ski}. At such low-$Q^2$ 
processes, the predictive power of HB$\chi$PT becomes very effective. Furthermore, the power counting 
of HB$\chi$PT allows a systematic control of the uncertainties involved. These uncertainties could be 
improved order-by-order in the expansion scheme of HB$\chi$PT. This non-relativistic field theory 
has been used extensively in the past to study the physical properties and the low-energy dynamics of 
nucleons and other baryons~\cite{Bernard:1995dp,Bernard:2007}. In this work we use the same framework 
to analytically evaluate the TPE {\it box} and {\it cross box} diagrams (called {\it TPE box} hereafter) 
and the so-called {\it seagull} diagram (c.f. Fig.~\ref{tpe}.) While all the TPE box diagrams are 
ultraviolet (UV) finite, the seagull diagram is both IR as well as UV finite. In this work we shall use 
the gauge invariance-preserving dimensional regularization (DR) scheme in order to remove the IR 
singularities from the TPE box diagrams. To the best of our knowledge such a TPE evaluation in the 
context of HB$\chi$PT has not been pursued till date.  

The paper is organized as follows. In Sec.~\ref{theory} we introduce the general formalism for elastic 
lepton-proton scattering within HB$\chi$PT, providing the relevant terms from the chiral Lagrangian, 
{\it up-to-and-including} NLO in the chiral power counting, that is necessary for the evaluation of our 
TPE diagrams. We also discuss some of the details of the kinematics involved in the calculations, which 
are commensurate with the proposed MUSE kinematic domain. In Sec.~\ref{TPE}, we outline the crucial steps 
involved in the systematic removal of the IR divergences from the TPE diagrams at $\mathcal{O}(\alpha^3)$. 
Especially, we  discuss the subtle nature of many cancellations among the NLO TPE amplitudes in the soft 
photon limit and their relation to the corresponding soft photon bremsstrahlung processes. Next in 
Sec.~\ref{results} we present our numerical estimates of the TPE contribution to the unpolarized elastic 
cross section. Finally in Sec.~\ref{conclusion} we draw some conclusions and present our outlook. An appendix 
is included at the end which collects some of the details of our analytical evaluation of the seagull diagram.     

\section{Heavy Baryon Chiral Perturbation Theory treatment of Lepton-Proton Scattering}
\label{theory}
The relevant parts of the LO and NLO chiral Lagrangian needed in our TPE evaluation of 
the lepton proton scattering amplitudes are given in, e.g., Ref.~\cite{Bernard:1995dp}.  
Since at NLO the TPE vertices do not involve  pions, we ignore the pion degrees of freedom 
in the part of the chiral Lagrangian that we use (the pion loops arise at 
{\it next-to-next-to-leading order} (NNLO) which is beyond the accuracy of this work).    
From Ref.~\cite{Bernard:1995dp} we obtain 
\begin{equation}
\mathcal{L}_N = \mathcal{L}_{N}^{(\nu=0)} + \mathcal{L}_{N}^{(\nu=1)} + \cdots \,,
\end{equation}
where, the chiral indices $\nu=0$ and $\nu=1$ represent the  LO and NLO components 
of the HB$\chi$PT Lagrangian\footnote{Ideally $M$ 
is the mass of the nucleon in the chiral limit. In this work we use $M$ also to denote the proton's 
physical mass, $M=938.28$ MeV.}   
\begin{eqnarray}
\mathcal{L}_{N}^{(0)} &=& \bar N (i v\cdot D + g_A \, S\cdot u )N\,, 
\\
\mathcal{L}_{\pi N}^{(1)} &=& \bar N \left\{\frac{1}{2 M}(v\cdot D)^2 - \frac{1}{2 M}D\cdot D + \cdots \right\}N\,.
\end{eqnarray}
Here $N=(\rm{p\,\,n})^{\rm T}$ is the heavy nucleon spin-isospin field, and $v_\mu$ and $S_\mu$ are the 
nucleon velocity and spin four-vectors satisfying the condition, $v\cdot S=0$. Here we choose 
$v=(1,{\bf 0})$ such that $S=(0,\boldsymbol{\sigma}/2)$. The covariant derivatives in the Lagrangian are 
\begin{eqnarray}
D_\mu &=& \partial_\mu + \Gamma_\mu - i v_\mu^{(s)}\,,  \qquad u_\mu = i u^\dagger \nabla_\mu U u^\dagger\,, \\
\Gamma_\mu &=& \frac{1}{2}[u^\dagger(\partial_\mu-i r_\mu)u+u(\partial_\mu-i l_\mu)u^\dagger]\,,\\
\nabla_\mu U &=& \partial_\mu U-i r_\mu U+i U l_\mu\,.
\end{eqnarray}
Due to the absence of the explicit pions the $u = \sqrt{U}$ field is simply $u=I\equiv I_{2\times 2}$, the 
identity matrix in isospin space. The external iso-vector right- and left-fields, $r_\mu$ and $l_\mu$, 
respectively, have in our case simple expressions since the only external source field is the photon 
field $A_\mu(x)$. The chiral Lagrangian is therefore reduced to a combination of external iso-scalar 
source $v_\mu^{(s)}=-e\frac{I}{2}A_\mu(x)$ and iso-vector source $u_\mu=(l_\mu+r_\mu)/2=-e\frac{\tau^3}{2}A_\mu(x)$, 
where $\tau^3$ is the third Pauli isospin matrix. \\   

\begin{figure}[bp]
\centering
\includegraphics[scale=0.47]{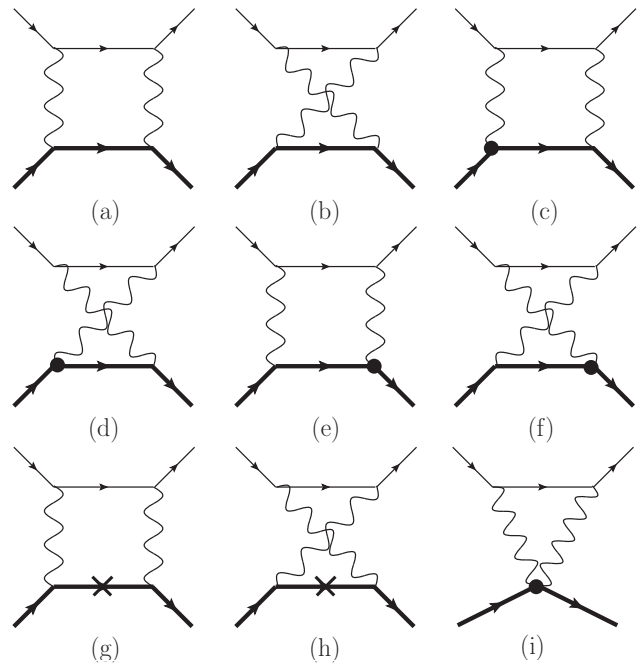}
\caption{\label{tpe} The TPE Feynman diagrams of $\mathcal{O}(e^4)$ which contribute to the 
        $\mathcal{O}(\alpha^3)$ interference term in the elastic lepton-proton cross section. 
        Thin lines represent lepton propagators, thick lines represent proton propagators, and 
        wiggly lines represent photon propagators. The solid dark circles and the lines with a 
        cross represent vertex and proton propagator insertions, respectively, from the NLO 
        Lagrangian $\mathcal{L}^{(1)}_{\pi N}$. Diagrams (a)-(h) are the ``box'' and ``cross-box'' 
        terms, and diagram (i) is the ``seagull'' term.}
\end{figure}
	
The relevant TPE amplitudes of $\mathcal{O}(e^4)$ are diagrammatically illustrated in  Fig.~\ref{tpe}. In the 
diagrams labeled (a) and (b), the proton-photon vertices arise from the LO chiral Lagrangian, while  
diagrams (c)-(f) contain one proton-photon vertex insertion from the NLO Lagrangian. Each of the diagrams (g) and 
(h) contains an NLO propagator insertion, and finally the seagull diagram (i) contains no intermediate proton, 
instead this diagram has an effective two-photon interaction vertex associated with the proton originating from 
the NLO Lagrangian. 

\begin{figure}[tbp]
\centering
\includegraphics[scale=0.65]{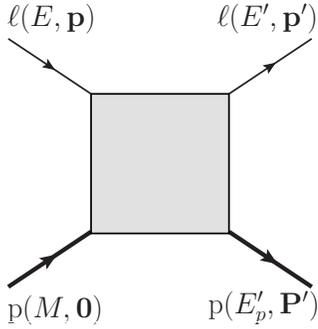}
\caption{\label{kinematics} The kinematics for the $\ell$p $\to\ell$p scattering 
        in the laboratory frame. The square shaded area represents all possible internal 
        graphs contributing to the elastic scattering process.}
\end{figure}

In this work we find it convenient to use the laboratory or rest frame of the proton target which 
allows a straightforward relation to the proposed MUSE kinematic. The convention used here is shown 
in Fig.~\ref{kinematics}, where the incoming lepton momentum $p = (E,\,{\bf p})$, the outgoing lepton 
momentum $p^\prime = (E^\prime,\,{\bf p^\prime})$, the incoming proton momentum $P = (M,\,{\bf 0})$ ;
and the outgoing proton momentum $P^\prime = (E_p^\prime,\,{\bf P^\prime})$. Additionally, in the HB$\chi$PT 
formalism one introduces a small so-called {\it residual} incoming proton momentum $p_p$ as defined through 
the relation $P^\mu = M v^\mu+p_p^\mu$ with $p_p^2\ll M^2$, which in the laboratory frame means $v\cdot p_p=0$. 
Similarly, the small residual outgoing proton momentum $p_p^\prime$ defined as 
${P^\prime}^\mu = M v^\mu+{p^{\prime}_p}^\mu$, where $(p_p^\prime )^2\ll M^2$, implies
\beq
v\cdot p^{\prime}_p = \frac{{\bf p^\prime_p}^{2}}{2M}+\mathcal{O}(M^{-2})\,.
\label{eq:v.p}
\eeq 
Finally, the four-momentum transfer in the elastic process is 
$Q_\mu = p_\mu-p_\mu^\prime=P^\prime_\mu-P_\mu=(p^\prime_p)_\mu-(p_p)_\mu$, and the lepton scattering angle is $\theta$.  

The MUSE collaboration has chosen the incident lepton momenta to have the following values: $115$ MeV/c, $153$ MeV/c 
and $210$ MeV/c. This means that for elastic scattering the four-momentum transfer $Q^2$ depends only on the scattering 
angle $\theta$. The corresponding range of the $Q^2$ value in the laboratory frame can be obtained using the relations
\beq
Q^2=2m_l^2-2E E^\prime(1-\beta\beta^\prime\cos\,\theta)=-2M(E-E^\prime)\,, 
\label{eq:Q^2}
\eeq 
where $\beta=|{\bf p}|/E$ and $\beta^\prime=|{\bf p^\prime}|/E^\prime$ are the incoming and outgoing lepton velocities, 
respectively. It may be shown that $0<|Q^2|< |Q^2_{\rm max}| =4M^2(E^2-m^2_l)/(m^2_l+M^2+2ME)$ 
represents the kinematically allowed (physical) range of momentum transfers~\cite{tomalak2014two}. However, the 
(laboratory frame) scattering angle is proposed by MUSE to be in the range 
$\theta\in [20^{\circ}, 100^{\circ}]$~\cite{Gilman:2013eiv}, for which the possible $Q^2$ range of values obtained from 
Eq.~\eqref{eq:Q^2} are tabulated in Table~\ref{q2table}. By examining the $Q^2$ values in the table we observe that 
$Q/\Lambda_\chi \ll 1$, i.e., the HB$\chi$PT power counting scheme can be applied reasonably well in the domain of the 
MUSE kinematics.\footnote{We note that for the TPE diagrams with either NLO vertex or propagator insertions being already 
of ${\mathcal O}({M^{-1})}$, it is reasonable to write $E^\prime=E+\frac{Q^2}{2M}\approx E$ at NLO.}  Here we 
remark that the lepton mass is explicitly included in all our expressions. In the next section we evaluate the TPE 
diagrams in  Fig.~\ref{tpe} and isolate the IR divergences of the TPE box diagrams.  

\begin{table}[bp]
\centering
\begin{tabular}{|c||c|c|c|}
\hline\hline
Momentum (p) in GeV/c & \,\,$0.115$  \,\,& \,\,$0.153$  \,\,& \,\,$0.210$ \,\,\\
\hline\hline
\multicolumn{4}{|c|}{$|Q^2|$ in (GeV/c)$^2$ for Electron}\\
\hline
Angle $\theta=20^{\circ}$ & $0.0016$ & $0.0028$ & $0.0052$\\
\hline
Angle $\theta=100^{\circ}$ & $0.027$ & $0.046$ & $0.082$\\
\hline
\multicolumn{4}{|c|}{$|Q^2|$ in (GeV/c)$^2$ for Muon}\\
\hline
Angle $\theta=20^{\circ}$ & $0.0016$ & $0.0028$ & $0.0052$\\
\hline
Angle $\theta=100^{\circ}$ & $0.026$ & $0.045$ & $0.080$\\
\hline\hline
\end{tabular}
\caption{\label{q2table} The MUSE range of $|Q^2|$ values for $e$p and $\mu$p scattering at the two 
        limits of the laboratory frame scattering angle, namely, $\theta=20^\circ$ and $100^\circ$, 
        obtained from Eq.~\eqref{eq:Q^2}. } 
\end{table}

\section{Two-Photon Exchange in the Soft Photon Approximation}    
\label{TPE}

In this section we  evaluate all the TPE diagrams in Fig.~\ref{tpe} using HB$\chi$PT and derive the $Q^2$ or 
$\theta$ dependence on the IR subtracted TPE diagrams in a gauge-invariant manner. The finite (IR subtracted) 
part of the TPE fractional corrections $\bar{\delta}_{\gamma\gamma}$ up to and including next-to-leading order 
accuracy, i.e., ${\mathcal O}(\alpha, M^{-1})$, to the elastic scattering cross section is defined by:
\begin{eqnarray}
\left[\frac{d\sigma_{el}(Q^2)}{d\Omega^{\,\prime}_l}\right]_{\gamma\gamma} = \left[\frac{d\sigma_{el}(Q^2)}{d\Omega^{\,\prime}_l}\right]_{\gamma}\!\!\bar{\delta}_{\gamma\gamma}(Q^2)\,, 
\label{tpe_crosssection}
\end{eqnarray}
where
\begin{eqnarray}
\bar{\delta}_{\gamma\gamma}(Q^2) = \frac{2\, {\mathcal R}e \sum_{spins} \left({\mathcal M}^{*}_\gamma \,{\mathcal M}_{\gamma\gamma}\right)}{\sum_{spins} |{\mathcal M}_\gamma|^2}-\delta^{\rm (box)}_{\rm IR}(Q^2) \,.\quad{} 
\label{tpe_delta}
\end{eqnarray} 
In this expression ${\mathcal M}_\gamma$  is the one-photon exchange (Born) amplitude,  
\begin{equation}
{\mathcal M}_\gamma=-\frac{e^2}{Q^2}\left[\bar u(p^\prime)\gamma^\mu u(p)\right]\,
\left[\chi^\dagger(p^\prime_p)v_\mu\chi(p_p)\right]\, ,
\label{eq:Mgamma}
\end{equation} 
and ${\mathcal M}_{\gamma\gamma}={\mathcal M}^{\rm (box)}_{\gamma\gamma}+{\mathcal M}^{(i)}_{\rm seagull}$ is the total TPE 
amplitude obtained by summing the TPE box amplitudes, ${\mathcal M}^{\rm (box)}_{\gamma\gamma}$, and the seagull amplitude, 
${\mathcal M}^{(i)}_{\rm seagull}$, viz, Feynman diagrams (a) - (i) in Fig.~\ref{tpe}. The corresponding Born cross 
section is $\left(d\sigma_{el}/d\Omega^{\,\prime}_l\right)_{\gamma}$, which to LO in the chiral expansion (including phase 
space $1/M$ proton recoil contributions) is given by
\begin{eqnarray}
\left[\frac{d\sigma_{el}(Q^2)}{d\Omega^{\,\prime}_l}\right]_{\gamma} 
&=& \frac{1}{64\pi^2M^2}\left(\frac{\beta^\prime E^\prime}{E}\right)\left[1+\frac{E}{M}(1-\cos\theta)\right]^{-1}
\nonumber\\
&& \times\, \frac{1}{4}\sum_{spins}\left|{\mathcal M}_{\gamma}\right|^2\,,
\nonumber\\
\frac{1}{4}\sum_{spins}\left|{\mathcal M}_{\gamma}\right|^2 
&=& \frac{64\pi^2\alpha^2}{Q^4}MEE^\prime(M+E^\prime_p)
\nonumber\\
&& \times \,\left[1+\beta\beta^\prime\cos\theta+\frac{m^2_l}{EE^\prime}\right]\,,
\end{eqnarray}
where the kinematics at $1/M$ order accuracy yield the following relations:
\begin{eqnarray}
E^\prime &=& E\left[1-\frac{\beta^2E}{M}(1-\cos\theta)+{\mathcal O}\left(M^{-2}\right)\right]\,,
\nonumber\\
\beta^\prime &=& \beta\left[1-\frac{(1-\beta^2)E}{M}(1-\cos\theta)+{\mathcal O}\left(M^{-2}\right)\right]\,,
\nonumber\\
Q^2 &=& -2\beta^2E^2(1-\cos\theta)
\nonumber\\
&& \times \,\left[1-\frac{E}{M}(1-\cos\theta)+{\mathcal O}\left(M^{-2}\right)\right]\,.
\end{eqnarray} 
In Eq.~\eqref{tpe_delta}, the term  ${\delta}^{\rm (box)}_{\rm IR}$ denotes the IR singular part of the TPE box diagrams'  
contribution to the elastic cross section. Utilizing DR we project out these singularities before deriving the 
expression for ${\delta}^{\rm (box)}_{\rm IR}$. It will be shown in a future publication~\cite{Talukdar} that the 
${\cal O}(e^3)$ soft bremsstrahlung amplitude has an IR singularity which in the cross section generates a singular 
term, $\delta^{\rm (soft)}_{\rm IR}$, which cancels the IR singularity in Eq.~\eqref{tpe_delta}, namely, 
${\delta}^{\rm (soft)}_{\rm IR}=-{\delta}^{\rm (box)}_{\rm IR}$ at ${\cal O}(\alpha^3)$. Our calculations of the finite TPE 
contribution $\bar{\delta}_{\gamma\gamma}$ in Eq.~\eqref{tpe_delta} inherently rely on the widely used  
{\it soft photon approximation} (SPA). While the HB$\chi$PT evaluation details and discussion of the full QED radiative 
corrections to the $\ell$p elastic scattering at NLO will be presented in Ref.~\cite{Talukdar}, here we simply quote our 
analytical expression for $\delta^{\rm (soft)}_{\rm IR}$:
\begin{eqnarray}
\delta^{\rm (soft)}_{\rm IR}(Q^2)&=& 
\frac{\alpha Q^2}{2\pi M E}\left\{\frac{1}{\epsilon}-\gamma_E+\ln\left(\frac{4\pi\mu^2}{-Q^2}\right)\right\}
\nonumber\\
&& \!\!\!\! \times \left\{\frac{1}{\beta}\ln\sqrt{\frac{1+\beta}{1-\beta}}+\frac{E}{E^\prime\beta^\prime}\ln\sqrt{\frac{1+\beta^\prime}{1-\beta^\prime}}\right\}\, . \,\,\,\,\quad{}
\label{eq:delta_box}
\end{eqnarray}

The TPE diagrams in Fig.~\ref{tpe} naturally include the contributions to the Coulomb wave functions describing 
the incoming and outgoing charged leptons. For example, the so-called ``Coulomb focusing'' or distortion of the 
scattered lepton spectrum at low-$|Q^2|$ is explained by considering one of the exchanged photons in the box 
diagrams as a soft photon.\footnote{The Coulomb distortion of the outgoing electron waves in the second Born 
approximation was investigated in Refs.~\cite{Sick:1998cvq,Blunden:2005jv} (see also Figs.~13 and 17 of 
Ref.~\cite{Arrington:2011dn}) where it was found to significantly enhance the cross section at backward 
scattering angles. However, in this case the contribution to the proton's rms charge radius is unlikely to be 
affected from the extrapolation of the form factor data to extreme forward angles where Coulomb wave function 
effects are found to be small. Hence, the authors of Refs.~\cite{Sick:1998cvq,Blunden:2005jv} concluded that the 
radius discrepancy could not have been  attributed to an erroneous experimental measurement due to the influence 
of this kind of SPA in the TPE contributions.} The SPA has widely been used in the literature as a practical tool 
to isolate the IR singularities of the TPE box diagrams. However, the exact implementation of the SPA is somewhat 
\textit{ad hoc} and differs in different theoretical works. For example, following the work of Maximon and 
Tjon~\cite{Maximon:2000hm}, the SPA is used only in the denominator  (propagators) of the integrand in order 
to single out the IR-divergent TPE amplitude, i.e., the momentum of the soft exchange photon is set to zero. 
Maximon and Tjon do not set the photon momentum to zero in the numerator of the integrand. On the other hand, 
following the work of Mo and Tsai~\cite{Mo:1968cg}, the SPA is used simultaneously in the numerator and 
denominator. As was noted in Ref.~\cite{Maximon:2000hm}, the convenience of using the former ``less drastic'' 
type of approximation is that the resulting expressions become somewhat simpler. However, some authors, e.g., those of 
Ref.~\cite{Koshchii:2017dzr}, have argued in favor of the latter ``more drastic'' approximation being more 
self-consistent. The essential point is to let the momenta associated with the soft photon go to zero, 
irrespective of whether they appear in the numerator or the denominator. Since these soft momentum factors, 
which appear in the numerator of the amplitudes, originate from those  in the denominator, it seems somewhat 
unreasonable to let them go to zero {\it only} in the denominator. Concurring with the argument presented in 
Ref.~\cite{Koshchii:2017dzr}, in the following we shall use the SPA definition of Mo and Tsai~\cite{Mo:1968cg}.  

As shown in the Figs.~\ref{tpe}(a)-(i), the loop integrals up to NLO in HB$\chi$PT contributing to the TPE amplitude
${\mathcal M}_{\gamma\gamma}$ are, respectively, given by
\begin{widetext}
\begin{eqnarray}
i{\mathcal M}^{(a)}_{\rm box} 
&=&e^4\int \frac{d^4k}{(2\pi)^4}
\frac{\left[\bar u(p^\prime)\gamma^\mu(\slashed{p}-\slashed{k}+m_l)\gamma^\nu u(p)\right]\,
\left[\chi^\dagger(p_p^\prime)v_\mu v_\nu \chi(p_p)\right]}{(k^2+i0)\,[(Q-k)^2+i0]\,(k^2-2k\cdot p+i0)\,(v\cdot k+i0)}\,,
\label{M_a}
\end{eqnarray}
\begin{eqnarray}
i{\mathcal M}^{(b)}_{\rm xbox}
&=&e^4\int \frac{d^4k}{(2\pi)^4}
\frac{\left[\bar u(p^\prime)\gamma^\mu(\slashed{p}-\slashed{k}+m_l)\gamma^\nu u(p)\right]\,
\left[\chi^\dagger(p_p^\prime)v_\mu v_\nu \chi(p_p)\right]}{(k^2+i0)\,[(Q-k)^2+i0]\,(k^2-2k\cdot p+i0)\,(-v\cdot k+i0)}\,,
\label{M_b}
\\
\nonumber\\
\nonumber\\
i{\mathcal M}^{(c)}_{\rm box}
&=&\frac{e^4}{2M}\int 
\frac{d^4k}{(2\pi)^4}
\frac{\left[\bar u(p^\prime)\gamma^\mu(\slashed{p}-\slashed{k}+m_l)\gamma^\nu u(p)\right]\,
\left[\chi^\dagger(p_p^\prime)\{v_\mu(2p_p+k)_\nu-v_\mu v_\nu (v\cdot k)\} \chi(p_p)\right]}{(k^2+i0)\,[(Q-k)^2+i0]\,(k^2-2k\cdot p+i0)\,(v\cdot k +i0)}\,,
\label{M_c}
\\
\nonumber\\
\nonumber\\
i{\mathcal M}^{(d)}_{\rm xbox}
&=&\frac{e^4}{2M}\int \frac{d^4k}{(2\pi)^4}
\frac{\left[\bar u(p^\prime)\gamma^\mu(\slashed{p}-\slashed{k}+m_l)\gamma^\nu u(p)\right]\,
\left[\chi^\dagger(p_p^\prime)\{v_\nu(p_p+ p_p^\prime-k)_\mu-v_\mu v_\nu (-v\cdot k))\} \chi(p_p)\right]}{(k^2+i0)\,[(Q-k)^2+i0]\,(k^2-2k\cdot p+i0)\,(-v\cdot k +i0)}\,,
\label{M_d}
\\
\nonumber\\
\nonumber\\
i{\mathcal M}^{(e)}_{\rm box}
&=&\frac{e^4}{2M}\int \frac{d^4k}{(2\pi)^4}
\frac{\left[\bar u(p^\prime)\gamma^\mu(\slashed{p}-\slashed{k}+m_l)\gamma^\nu u(p)\right]\,
\left[\chi^\dagger(p_p^\prime)\{v_\nu(p_p+ p_p^\prime+k)_\mu-v_\mu v_\nu (v\cdot k)\} \chi(p_p)\right]}{(k^2+i0)\,[(Q-k)^2+i0]\,(k^2-2k\cdot p+i0)\,(v\cdot k +i0)}\,,
\label{M_e}
\\
\nonumber\\
\nonumber\\
i{\mathcal M}^{(f)}_{\rm xbox}
&=&\frac{e^4}{2M}\int \frac{d^4k}{(2\pi)^4}
\frac{\left[\bar u(p^\prime)\gamma^\mu(\slashed{p}-\slashed{k}+m_l)\gamma^\nu u(p)\right]\,
\left[\chi^\dagger(p_p^\prime)\{v_\mu(2p_p^\prime-k)_\nu-v_\mu v_\nu (-v\cdot k))\} \chi(p_p)\right]}{(k^2+i0)\,[(Q-k)^2+i0]\,(k^2-2k\cdot p+i0)\,(-v\cdot k +i0)}\,,
\label{M_f}
\\
\nonumber\\
\nonumber\\
i\mathcal{M}^{(g)}_{\rm box}
&=&\frac{e^4}{2M}\int \frac{d^4k}{(2\pi)^4}
\frac{\left[\bar u(p^\prime)\gamma^\mu(\slashed{p}-\slashed{k}+m_l)\gamma^\nu u(p)\right]\,
\left[\chi^\dagger(p_p^\prime)v_\mu v_\nu\chi(p_p)\right]}{(k^2+i0)\,[(Q-k)^2+i0]\,(k^2-2k\cdot p+i0)}
\left(1+\frac{p^2_p}{(v\cdot k)^2}-\frac{(p_p+k)^2}{(v\cdot k)^2}\right)\,,
\label{M_g}
\\
\nonumber\\
\nonumber\\
i{\mathcal M}^{(h)}_{\rm xbox}
&=&\frac{e^4}{2M}\int \frac{d^4k}{(2\pi)^4}
\frac{\left[\bar u(p^\prime)\gamma^\mu(\slashed{p}-\slashed{k}+m_l)\gamma^\nu u(p)\right]\,
\left[\chi^\dagger(p_p^\prime)v_\mu v_\nu\chi(p_p)\right]}{(k^2+i0)\,[(Q-k)^2+i0]\,(k^2-2k\cdot p+i0)}
\left(1+\frac{p^{\prime\,2}_p}{(v\cdot k)^2}-\frac{(p_p^\prime-k)^2}{(v\cdot k)^2}\right)\,,
\label{M_h}
\\
\nonumber\\
\nonumber\\
i{\mathcal M}^{(i)}_{\rm seagull}
&=&\frac{2 e^4}{2M}\int \frac{d^4k}{(2\pi)^4}
\frac{\left[\bar u(p^\prime)\gamma^\mu(\slashed{p}-\slashed{k}+m_l)\gamma^\nu u(p)\right]\,
\left[\chi^\dagger(p_p^\prime)(v_\mu v_\nu-g_{\mu\nu})\chi(p_p)\right]}{(k^2+i0)\,[(Q-k)^2+i0]\,(k^2-2k\cdot p+i0)}\,,
\label{M_i}
\end{eqnarray}
\end{widetext}
where $u(p)$ and $\bar{u}(p^\prime)$ are the incoming and outgoing lepton Dirac spinors, and, $\chi(p)$ and 
$\chi^\dagger(p^\prime_p)$ are the proton's non-relativistic two-component Pauli spinors. Here we remark that 
the two LO TPE integrals, Eqs.~(\ref{M_a}) and (\ref{M_b}), should contain the kinetic energy terms of 
${\cal O}(M^{-1})$ in the proton propagators [c.f. Eq.~\eqref{eq:v.p}]. However, for the purpose of 
distinguishing the ``true'' LO from the NLO parts of the integrals, these ${\cal O}(M^{-1})$ terms from the 
two LO amplitudes have been included in the NLO propagator insertion integrals, Eqs.~(\ref{M_g}) and 
(\ref{M_h}), respectively. In that case the first two amplitudes, namely, $\mathcal{M}^{(a)}_{\rm box}$ and 
$\mathcal{M}^{(b)}_{\rm xbox}$, correspond to the ``true'' LO contribution of the TPE. All the rest 
contributing at NLO either correspond to terms that directly arise in the NLO chiral power counting or 
are attributed to the dynamical recoil $\mathcal O(M^{-1})$ terms moved from the LO chiral counting amplitudes. 
We isolate the IR divergences by taking the soft photon limit, which means: when one of the two photons' 
four-momenta is considered {\it soft} (either setting $k=0$ or $k=Q$) the other photon is {\it hard} (either 
with $(Q-k)^2\neq 0$ or $k^2\neq 0$). To project out the IR singular terms, we must evaluate the loop integrals 
at both poles and then consider their sum. 

To demonstrate the utility of this approach, let us apply SPA to the LO box, Eq.~\eqref{M_a}, and cross box, 
Eq.~\eqref{M_b}, amplitudes [c.f. Figs.~\ref{tpe}(a) and \ref{tpe}(b)]. Treating each of the two photons to be separately 
soft leads to the following sum of the amplitudes:  
\begin{widetext}
\begin{eqnarray}
i{\mathcal M}^{(a)}_{\rm box}
&\stackrel{\gamma_{\rm soft}}{\leadsto}&-\,2e^2\,(v\cdot p)\,{\mathcal M}_\gamma\int \frac{d^4k}{(2\pi)^4}
\frac{1}{k^2\,(k^2-2k\cdot p)\, (v\cdot k +i0)} 
\nonumber\\
 &&-\,2e^2\,(v\cdot p^\prime)\,{\mathcal M}_\gamma\int \frac{d^4k}{(2\pi)^4}
\frac{1}{k^2\,(k^2-2k\cdot p^\prime)\,(v\cdot k +i0)}\,,
\end{eqnarray}
and,
\begin{eqnarray}
i{\mathcal M}^{(b)}_{\rm xbox}
&\stackrel{\gamma_{\rm soft}}{\leadsto}& 2e^2\,(v\cdot p)\,{\mathcal M}_\gamma\int \frac{d^4k}{(2\pi)^4}
\frac{1}{k^2\,(k^2-2k\cdot p)\,(v\cdot k -i0)}
\nonumber\\
&&+\,2e^2\,(v\cdot p^\prime)\,{\mathcal M}_\gamma\int \frac{d^4k}{(2\pi)^4}
\frac{1}{k^2\,(k^2-2k\cdot p^\prime)\,(v\cdot k -i0)}
\end{eqnarray}
\end{widetext} 
It is immediately clear that $\mathcal{M}^{(a)}_{\rm box}$ is effectively canceled by  
$\mathcal{M}^{(b)}_{\rm xbox}$, which apparently is not manifest otherwise.\!\!\!~\footnote{Note 
the difference in the signs of the $\pm i\eta\to \pm i0$ terms in the heavy proton propagators 
of the two amplitudes contributes to a residual imaginary part, which is, however, irrelevant 
in the present context of evaluation of the unpolarized cross section.} Thus, we conclude that 
using the SPA  the LO amplitudes of the TPE diagrams give no LO amplitude contributions in 
HB$\chi$PT. This LO cancellation is anticipated since the proton does not generate any LO 
bremsstrahlung in HB$\chi$PT, {\em vis-a-vis} no LO IR divergence contributions in 
$\delta_{\rm IR}^{\rm (soft)}$~\cite{Talukdar}. A similar conclusion was obtained in 
Refs.~\cite{Talukdar:2018hia,Myhrer:2018ski}, which evaluated the lepton-proton bremsstrahlung 
process ($\ell$p $\to\ell$p$\gamma$) using the same HB$\chi$PT framework.  

The first non-vanishing TPE contributions in SPA arise from the NLO proton recoil contributions, 
which is commensurate with the corresponding non-zero HB$\chi$PT bremsstrahlung amplitudes with 
the soft photons radiated from NLO proton-photon vertices. We now analyze the NLO TPE integrals, 
Eqs.~\eqref{M_c}-\eqref{M_i}, and the following observations are in order\,:    
\begin{itemize}
\item First, when we sum these amplitudes, the terms containing the $v_\mu v_\nu$  in the amplitudes 
      ${\mathcal M}^{(c)}_{\rm box}$ through ${\mathcal M}^{(f)}_{\rm box}$ cancel with the first terms of 
      amplitudes ${\mathcal M}^{(g)}_{\rm box}$ and ${\mathcal M}^{(h)}_{\rm xbox}$ plus the $v_\mu v_\nu$  
      part of the seagull term ${\mathcal M}^{(i)}_{\rm seagull}$. 
\item Second, applying SPA to the remaining parts of the two amplitudes, $\mathcal{M}^{(g)}_{\rm box}$ 
      and $\mathcal{M}^{(h)}_{\rm xbox}$, we observe that they also cancel in the soft photon limits. 
      This is easily seen by analyzing the remaining parts of the NLO proton propagator in each of 
      these integrals in the following way:
\begin{widetext} 
\begin{eqnarray}
i\left[{\mathcal M}^{(g)}_{\rm box}+{\mathcal M}^{(h)}_{\rm xbox}\right]_{\rm residual} 
&=& \frac{e^4}{2M} \int\frac{d^4k}{(2\pi)^4} \cdots 
\left(\frac{p^2_p}{(v\cdot k)^2}-\frac{(p_p+k)^2}{(v\cdot k)^2}+\frac{p^{\prime\,2}_p}{(v\cdot k)^2}-\frac{(p_p^\prime-k)^2}{(v\cdot k)^2}\right)
\nonumber\\
&\stackrel{\gamma_{\rm soft}}{\leadsto}& \frac{e^4}{2M}\int\frac{d^4k}{(2\pi)^4} \cdots 
\left(\cancel{\frac{p^2_p}{(v\cdot k)^2}}-\cancel{\frac{p^2_p}{(v\cdot k)^2}}+\bcancel{\frac{p^{\prime\,2}_p}{(v\cdot k)^2}}-\bcancel{\frac{p^{\prime\,2}_p}{(v\cdot k)^2}}\right)_{k\to 0}
\nonumber\\
&&+\,\frac{e^4}{2M} \int\frac{d^4k}{(2\pi)^4} \cdots 
\left(\cancel{\frac{p^2_p}{(v\cdot k)^2}}-\bcancel{\frac{p^{\prime\,2}_p}{(v\cdot k)^2}}+\bcancel{\frac{p^{\prime\,2}_p}{(v\cdot k)^2}}-\cancel{\frac{p^2_p}{(v\cdot k)^2}}\right)_{k\to Q} = 0 \,.
\label{eq:gandh}
\end{eqnarray} 
\end{widetext} 
\end{itemize}  
Consequently, applying SPA, the two amplitudes, $\mathcal{M}^{(g)}_{\rm box}$ and $\mathcal{M}^{(h)}_{\rm xbox}$,  
which are the amplitudes with intermediate NLO proton propagator insertions, effectively do not contribute 
to the sum of the NLO TPE box amplitudes. Furthermore, we observe that in the seagull amplitude, Eq.~\eqref{M_i}, 
only the terms proportional to $g_{\mu\nu}$ contribute to the total TPE amplitude. The residual parts of the TPE 
integrals at NLO, after applying SPA to the box diagrams, yield  the following simplified soft photon 
amplitudes.\!\!~\footnote{We do not apply SPA in the evaluation of the IR-finite seagull diagram. 
A {\it naive} application of SPA to this diagram leads to lepton self-energy-like contributions with 
spurious IR-divergent terms. However, an exact evaluation (see Appendix) shows no such singularities.}  
\begin{widetext}
\begin{eqnarray}
i\widetilde{\mathcal M}^{(c)}_{\rm box}
&=&\frac{e^4}{M}(v\cdot p^\prime)\int \frac{d^4k}{(2\pi)^4}
\frac{\left[\bar u(p^\prime)\gamma^\nu u(p)\right]\,
\left[\chi^\dagger(p_p^\prime)(p_p+p^\prime_p)_\nu \chi(p_p)\right]}{k^2\,Q^2\,(k^2-2k\cdot p^\prime)\,(v\cdot k +i0)}
\nonumber\\
&&-\,\frac{2e^2}{M}(p\cdot p_p)\,{\mathcal M}_\gamma\int \frac{d^4k}{(2\pi)^4}
\frac{1}{k^2\,(k^2-2k\cdot p+i0)\,(v\cdot k +i0)} \,,
\end{eqnarray}
\begin{eqnarray} 
i\widetilde{\mathcal M}^{(d)}_{\rm xbox} 
&=&-\frac{e^4}{M}(v\cdot p)\int \frac{d^4k}{(2\pi)^4}
\frac{\left[\bar u(p^\prime)\gamma^\mu u(p)\right]\,
\left[\chi^\dagger(p_p^\prime)(p_p+p^\prime_p)_\mu \chi(p_p)\right]}{k^2\,Q^2\,(k^2-2k\cdot p)\,(v\cdot k -i0)}
\nonumber\\
&&+\,\frac{2e^2}{M}(p^\prime\cdot p_p)\,{\mathcal M}_\gamma\int \frac{d^4k}{(2\pi)^4}
\frac{1}{k^2\,(k^2-2k\cdot p^\prime)\,(v\cdot k -i0)}\,,
\\
\nonumber\\
\nonumber\\
i\widetilde{\mathcal M}^{(e)}_{\rm box}
&=&\frac{e^4}{M}(v\cdot p)\int \frac{d^4k}{(2\pi)^4}
\frac{\left[\bar u(p^\prime)\gamma^\mu u(p)\right]\,
\left[\chi^\dagger(p_p^\prime)(p_p+p^\prime_p)_\mu \chi(p_p)\right]}{k^2\,Q^2\,(k^2-2k\cdot p)\,(v\cdot k +i0)}
\nonumber\\
&&-\,\frac{2e^2}{M}(p^\prime\cdot p_p^\prime)\,{\mathcal M}_\gamma\int \frac{d^4k}{(2\pi)^4}
\frac{1}{k^2\,(k^2-2k\cdot p^\prime)\,(v\cdot k +i0)}\,,
\\
\nonumber\\
\nonumber\\
i\widetilde{\mathcal M}^{(f)}_{\rm xbox}
&=&-\,\frac{e^4}{M}(v\cdot p^\prime) \int \frac{d^4k}{(2\pi)^4}
\frac{\left[\bar u(p^\prime)\gamma^\nu u(p)\right]\,
\left[\chi^\dagger(p_p^\prime)(p_p+p^\prime_p)_\nu \chi(p_p)\right]}{k^2\,Q^2\,(k^2-2k\cdot p^\prime)\,(v\cdot k -i0)}
\nonumber\\
&&+\,\frac{2e^2}{M}(p\cdot p^\prime_p)\,{\mathcal M}_\gamma\int \frac{d^4k}{(2\pi)^4}
\frac{1}{k^2\,(k^2-2k\cdot p)\,(v\cdot k -i0)}\,,
\\
\nonumber\\
\nonumber\\
i\widetilde{\mathcal M}^{(i)}_{\rm seagull}
&=&-\,\frac{e^4}{M}\int \frac{d^4k}{(2\pi)^4}
\frac{\left[\bar u(p^\prime)\gamma^\mu(\slashed{p}-\slashed{k}+m_l)\gamma_\mu u(p)\right]\,
\left[\chi^\dagger(p_p^\prime)\chi(p_p)\right]}{k^2\,(Q-k)^2\,(k^2-2k\cdot p+i0)}\,,
\end{eqnarray}
\end{widetext}  
where the {\it tilde} symbols denote the residual NLO TPE amplitudes of Eqs.~\eqref{M_c}-\eqref{M_i}.     
Here we note again that there is a cancellation between $\widetilde{\mathcal M}^{(c)}_{\rm box}$ and 
$\widetilde{\mathcal M}^{(f)}_{\rm xbox}$ for the coefficient of $v\cdot p^\prime$, and between 
$\widetilde{\mathcal M}^{(d)}_{\rm xbox}$ and $\widetilde{\mathcal M}^{(e)}_{\rm box}$ for the coefficient of 
$v\cdot p$ (up to an irrelevant imaginary part.) Then the resulting sum of the NLO TPE box amplitudes 
in the soft photon limit gets ``factorized'' into a $Q^2$ dependent function $f(Q^2)$ times the Born 
amplitude ${\mathcal M}_{\gamma}$~\cite{BlundenPRL}, namely,

\vspace{-0.3cm}

\begin{widetext} 
\begin{eqnarray}
{\mathcal M}^{\rm (box)}_{\gamma\gamma}
\stackrel{\gamma_{\rm soft}}{\leadsto}\widetilde{\mathcal M}^{\rm (box)}_{\gamma\gamma} \equiv f(Q^2){\mathcal M}_{\gamma}
= \widetilde{\mathcal M}^{(c)}_{\rm box}+\widetilde{\mathcal M}^{(d)}_{\rm xbox}
+\widetilde{\mathcal M}^{(e)}_{\rm box}+\widetilde{\mathcal M}^{(f)}_{\rm xbox}\,,
\label{eq:Mbox1} 
\end{eqnarray}
where
\begin{eqnarray}
f(Q^2)=-\,\frac{2e^2}{M}\Big[(p\cdot p_p){\mathcal K}^{(+)}_v(p) - (p^\prime\cdot p_p){\mathcal K}^{(-)}_v(p^\prime) 
+ (p^\prime\cdot p^\prime_p){\mathcal K}^{(+)}_v(p^\prime) - (p\cdot p^\prime_p){\mathcal K}^{(-)}_v(p)\Big] \, . 
\label{eq:Mbox2} 
\end{eqnarray}
\end{widetext} 
The integrals ${\cal K}^{(+)}_v$ and ${\cal K}^{(-)}_v$ are solved in $D$ dimensions, i.e., $D>4$ is the 
analytically continued space-time dimension. In the expressions below, $\epsilon=(4-D)/2$, $\mu$ corresponds 
to the  subtraction scale, $\gamma_E = 0.577216...$ is the Euler-Mascheroni constant, and the integrals 
${\mathcal K}^{(\pm)}_v(p)$ 
in the above expression are given by 
\begin{widetext} 
\begin{eqnarray}
{\mathcal K}^{(+)}_v(p)
&=&\frac{1}{i}\int \frac{d^4k}{(2\pi)^4}\frac{1}{k^2(k^2-2k\cdot p)(v\cdot k +i0)}
\nonumber\\
&=&-\frac{1}{(4\pi)^2 E\beta}\left[\left\{\frac{1}{\epsilon}-\gamma_E +\ln\left(\frac{4\pi\mu^2}{m^2_l}\right)\right\}\ln\sqrt{\frac{1+
\beta}{1-\beta}}-\ln^2\sqrt{\frac{1+\beta}{1-\beta}}-\text{Sp}\left(\frac{2\beta}{1+\beta}\right)+\frac{\pi^2}{2}\right.
\nonumber\\
&&\left.\hspace{2cm}-i\pi\left\{\frac{1}{\epsilon}-\gamma_E+\ln\left(\frac{\pi\mu^2}{E^2\beta^2}\right)\right\}\,\right]\,,
\nonumber\\
\nonumber\\
{\mathcal K}^{(-)}_v(p)
&=&\frac{1}{i}\int \frac{d^4k}{(2\pi)^4}\frac{1}{k^2(k^2-2k\cdot p)(v\cdot k -i0)}
\nonumber\\
&=&-\frac{1}{(4\pi)^2 E\beta}\left[\left\{\frac{1}{\epsilon}-\gamma_E+\ln\left(\frac{4\pi\mu^2}{m^2_l}\right)\right\}\ln\sqrt{\frac{1+
\beta}{1-\beta}}-\ln^2\sqrt{\frac{1+\beta}{1-\beta}}-\text{Sp}\left(\frac{2\beta}{1+\beta}\right)\right]\,,  
\label{eq:Mbox3} 
\end{eqnarray}
\end{widetext} 
where the term `Sp' stands for the standard {\it Spence function} defined as the integral
\begin{equation}
{\rm Sp}(z)=\int^z_0 dt \frac{\ln(1-t)}{t}\,\,;\,\, z\in \mathbb{R}\,. 
\label{eq:Spence}
\end{equation}  
Likewise, we find the expression for the integrals ${\mathcal K}^{(\pm)}_v(p^\prime)$ in Eq.~(\ref{eq:Mbox2}) 
by replacing  $E\leftrightarrow E^\prime$ and $\beta\leftrightarrow \beta^\prime$. The IR divergences correspond 
to the poles in the dimensionally regularized integrals in the limit $\epsilon=(4-D)/2 \to 0^-$. The appearance 
of an imaginary part in these integrals depends on the sign of the $\pm i\eta$ term as $\eta\to 0$ in the proton 
propagator. As mentioned, the imaginary parts are irrelevant in our present context of the unpolarized cross 
section analysis. Nevertheless, it might be interesting to note that an additional IR divergence arises in the 
imaginary part of ${\mathcal K}^{(+)}_v(p)$ which is of importance in a polarized cross section analysis. 

Finally, we sum the TPE amplitudes and compute their interference with the Born amplitude in order to determine 
their contribution to the elastic cross section, with the appropriate IR singular term subtracted as shown in 
Eq.~\eqref{tpe_delta}. To this end, the sum of the factorizable IR-divergent TPE box amplitudes with the 
non-factorizable IR-free seagull amplitude (evaluated in the Appendix) is given as 
\begin{eqnarray}
{\mathcal M}_{\gamma\gamma}=\widetilde{\mathcal M}^{\rm (box)}_{\gamma\gamma} + \widetilde{\mathcal M}^{(i)}_{\rm seagull}
=f(Q^2){\mathcal M}_{\gamma} + \widetilde{\mathcal M}^{(i)}_{\rm seagull} \,.\qquad
\end{eqnarray}  
Denoting the corresponding fractional TPE contributions to the elastic cross section as   
\begin{eqnarray}
\delta_{\gamma\gamma}(Q^2)= \delta^{\rm (box)}_{\gamma\gamma}(Q^2) + \delta^{\rm (seagull)}_{\gamma\gamma}(Q^2)\, ,
\label{eq:delta_2gamma}
\end{eqnarray}   
we obtain the following NLO expressions,\!\!~\footnote{Since the corrections in Eqs.~(\ref{eq:deltaYY_box}) 
and (\ref{eq:deltaYY_seagull}) originate at NLO, we use the LO expression for the four-momentum transferred, 
i.e.,  $Q^2\to Q^2_0=-2\beta^2E^2(1-\cos\theta)$.} noting that $E^\prime=E+{\mathcal O}(M^{-1})$ and 
$\beta^\prime=\beta+{\mathcal O} (M^{-1})$: 
\begin{widetext}
\begin{eqnarray}
\delta^{\rm (box)}_{\gamma\gamma} (Q^2)
&=&\frac{2{\mathcal R}e\sum_{spins}\left({\mathcal M}^*_\gamma\,\widetilde{\mathcal M}^{\rm (box)}_{\gamma\gamma}\right)}{\sum_{spins}|{\mathcal M}_\gamma|^2} = 2{\mathcal R}e [f(Q^2)]
\nonumber\\
&=&-\frac{\alpha Q^2}{2\pi M E}\left[\left\{\frac{1}{\epsilon}-\gamma_E+\ln\left(\frac{4\pi\mu^2}{m^2_l}\right)\right\}
\left\{\frac{1}{\beta}\ln\sqrt{\frac{1+\beta}{1-\beta}}+\frac{E}{E^\prime\beta^\prime}\ln\sqrt{\frac{1+\beta^\prime}{1-\beta^\prime}}\right\} 
\right. + 
\nonumber\\
&&\left.+\frac{1}{\beta}\left\{\frac{\pi^2}{2}-\ln^2\sqrt{\frac{1+\beta}{1-\beta}}-\text{Sp}\left(\frac{2\beta}{1+\beta}\right)\right\}+
\frac{E}{E^\prime \beta^\prime}\left\{\frac{\pi^2}{2}-\ln^2\sqrt{\frac{1+\beta^\prime}{1-\beta^\prime}} -
\text{Sp}\left(\frac{2\beta^\prime}{1+\beta^\prime}\right)\right\}\,\right]  
\nonumber\\
\nonumber\\
&=&\delta^{\rm (box)}_{\rm IR}(Q^2)-\frac{\alpha Q^2}{\pi M E\beta}\left[\frac{\pi^2}{2}+\ln\left(\frac{-Q^2}{m^2_l}\right)\ln\sqrt{\frac{1+\beta}{1-\beta}}-\ln^2\sqrt{\frac{1+\beta}{1-\beta}}-\text{Sp}\left(\frac{2\beta}{1+\beta}\right)\right] + {\mathcal O}\left(\frac{1}{M^2}\right)\,,
\label{eq:deltaYY_box}
\end{eqnarray}
for the TPE box contribution, with $\delta^{\rm (box)}_{\rm IR}(Q^2)=-\delta^{\rm (soft)}_{\rm IR}(Q^2)$ as given in Eq.~\eqref{eq:delta_box}, and the finite seagull contribution
\begin{eqnarray}
\delta^{\rm (seagull)}_{\gamma\gamma} (Q^2)&=& \frac{2{\mathcal R}e\sum_{spin}\left({\mathcal M}^*_\gamma\,\widetilde{\mathcal M}^{(i)}_{\rm seagull}\right)}{\sum_{spin}|{\mathcal M}_\gamma|^2}
\nonumber\\
&=& -\frac{2\alpha Q^2}{\pi ME}\left[\frac{E^2+EE^\prime}{Q^2+4EE^\prime}\right]\left({\mathcal I}_1(Q^2)+{\mathcal I}_2(Q^2)+\frac{Q^2}{m^2_l}\left[{\mathcal I}_3(Q^2)-{\mathcal I}_4(Q^2)\right]\right)
\nonumber\\
\nonumber\\
&=& -\frac{4\alpha Q^2}{\pi ME}\left[\frac{E^2}{Q^2+4E^2}\right]\left({\mathcal I}_1(Q^2)+{\mathcal I}_2(Q^2)+\frac{Q^2}{m^2_l}\left[{\mathcal I}_3(Q^2)-{\mathcal I}_4(Q^2)\right]\right) + {\mathcal O}\left(\frac{1}{M^2}\right)\, .
\label{eq:deltaYY_seagull}
\end{eqnarray}
\end{widetext} 
The integrals ${\mathcal I}_i$ ($i=1-4$) are presented in Appendix where we evaluate the seagull term.
We subsequently use Eq.~\eqref{tpe_delta} to obtain the finite TPE contribution in SPA. As mentioned, 
$\delta^{\rm (box)}_{\rm IR}$  cancels exactly with $\delta^{\rm (soft)}_{\rm IR}$ when we include the soft 
bremsstrahlung contribution to this order in QED.

\section{Results and Discussion}
\label{results}
%
\begin{figure*}[tbp]
\centering
{\includegraphics[scale=0.48]{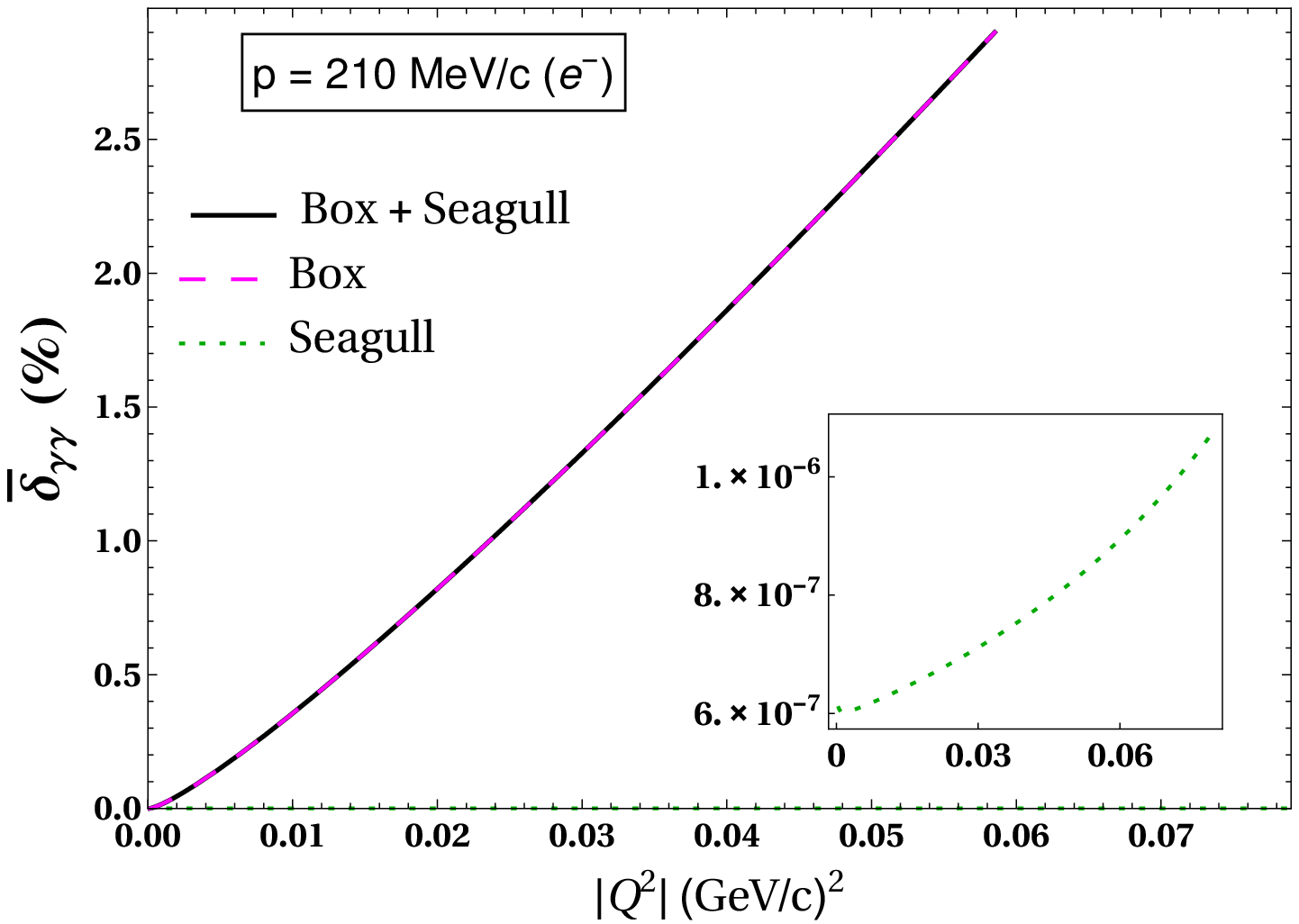}}\qquad\qquad{\includegraphics[scale=0.48]{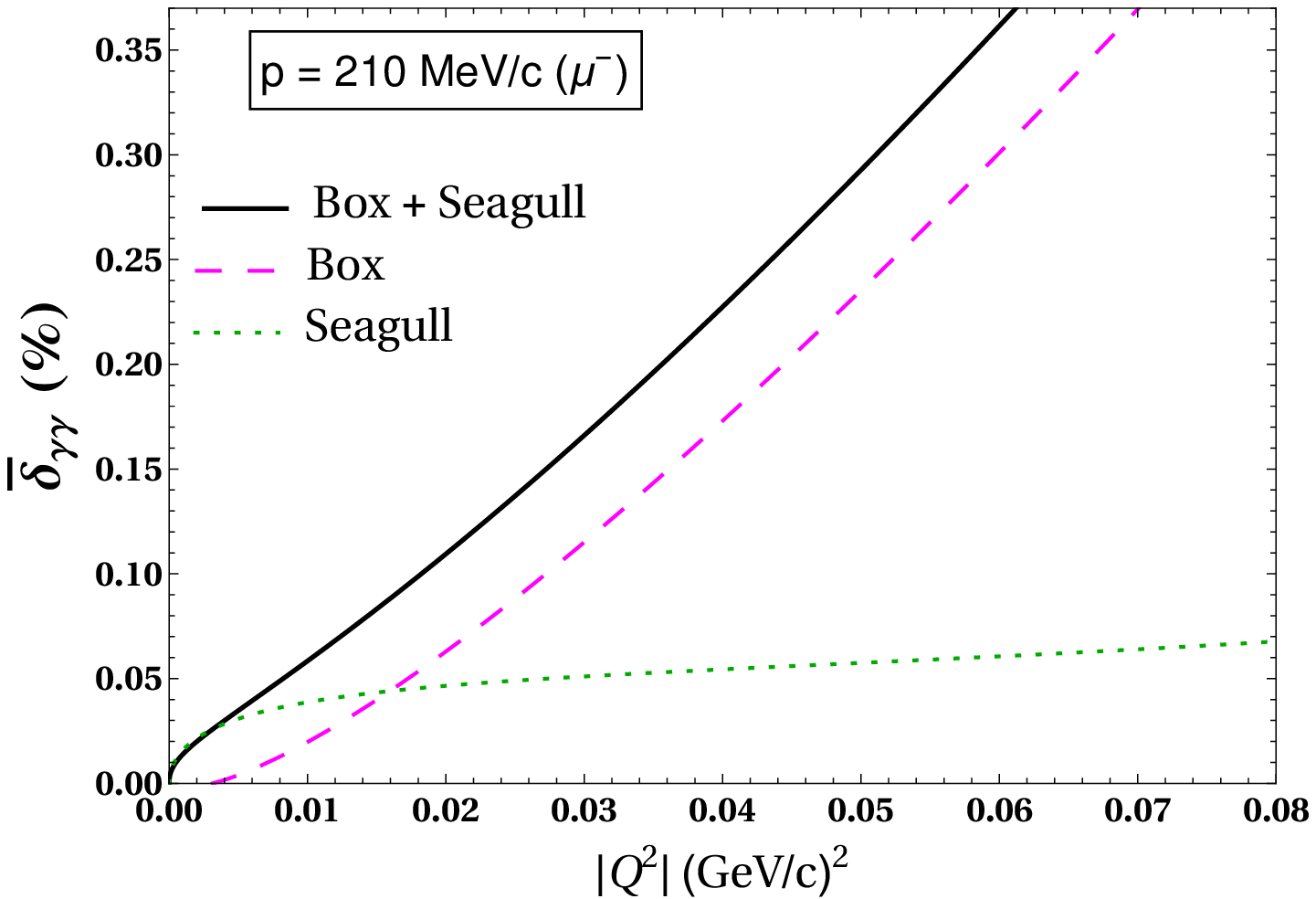}}

\vspace{0.4cm}

{\includegraphics[scale=0.48]{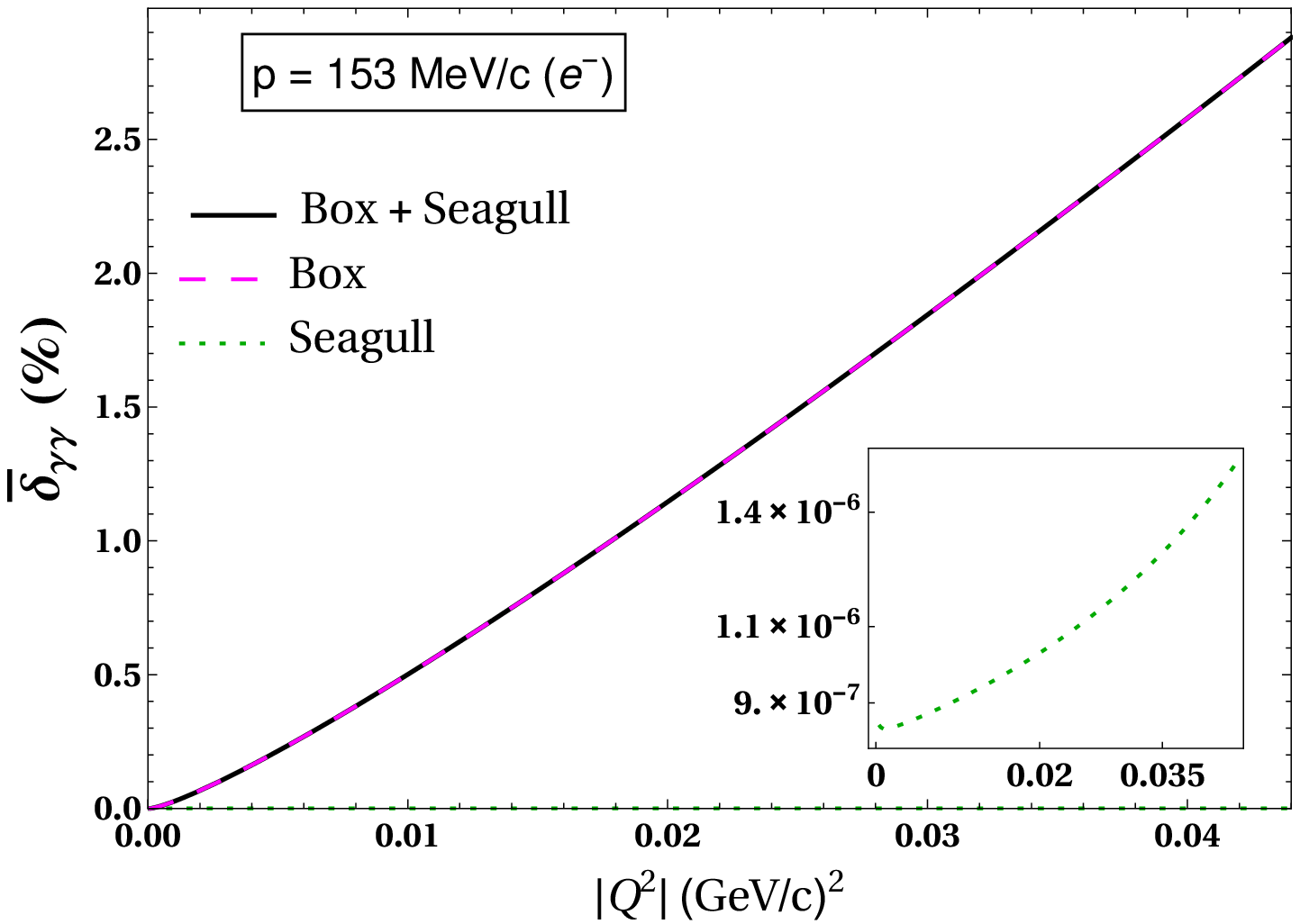}}\qquad\qquad{\includegraphics[scale=0.48]{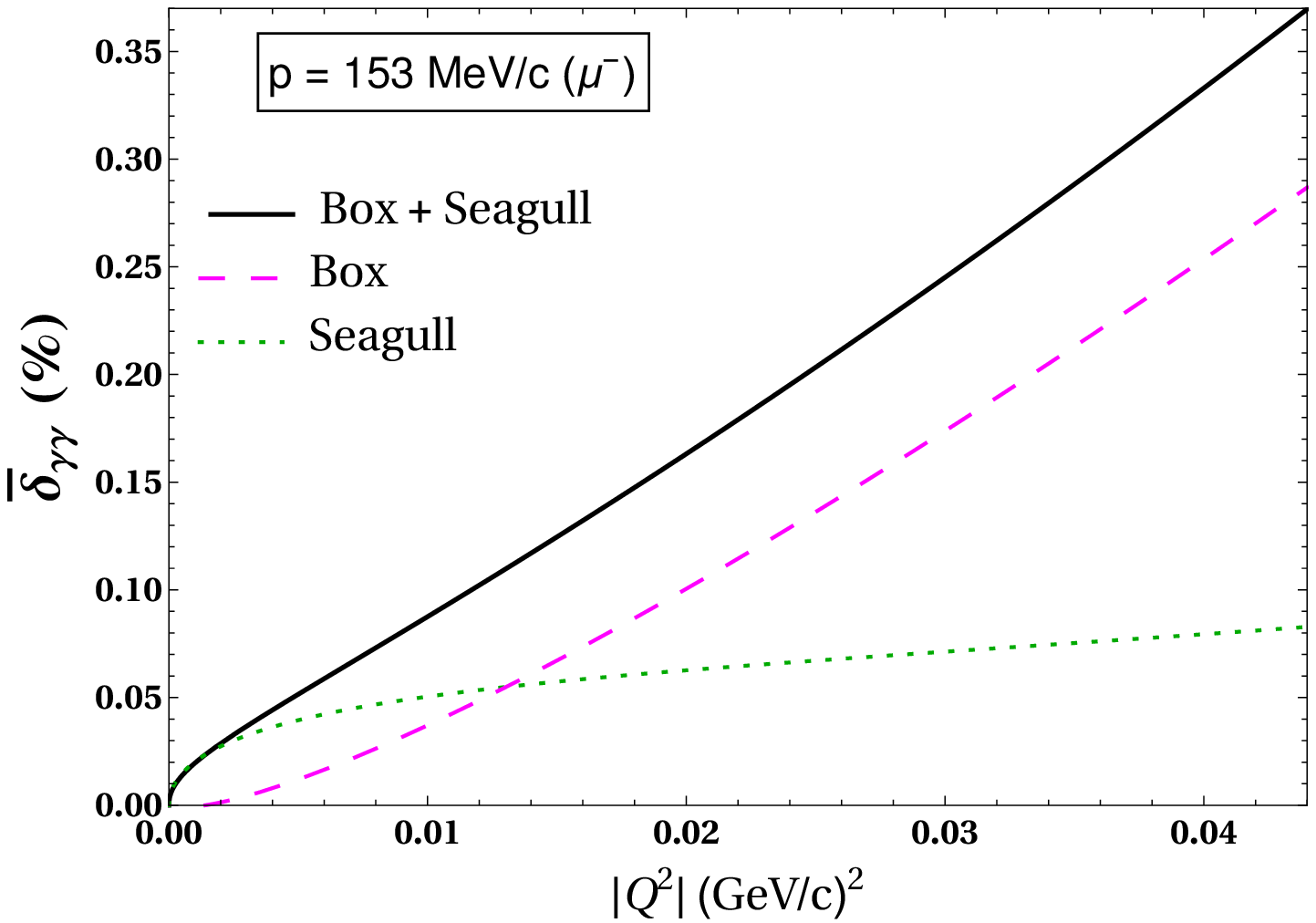}}\\ 

\vspace{0.4cm}

{\includegraphics[scale=0.48]{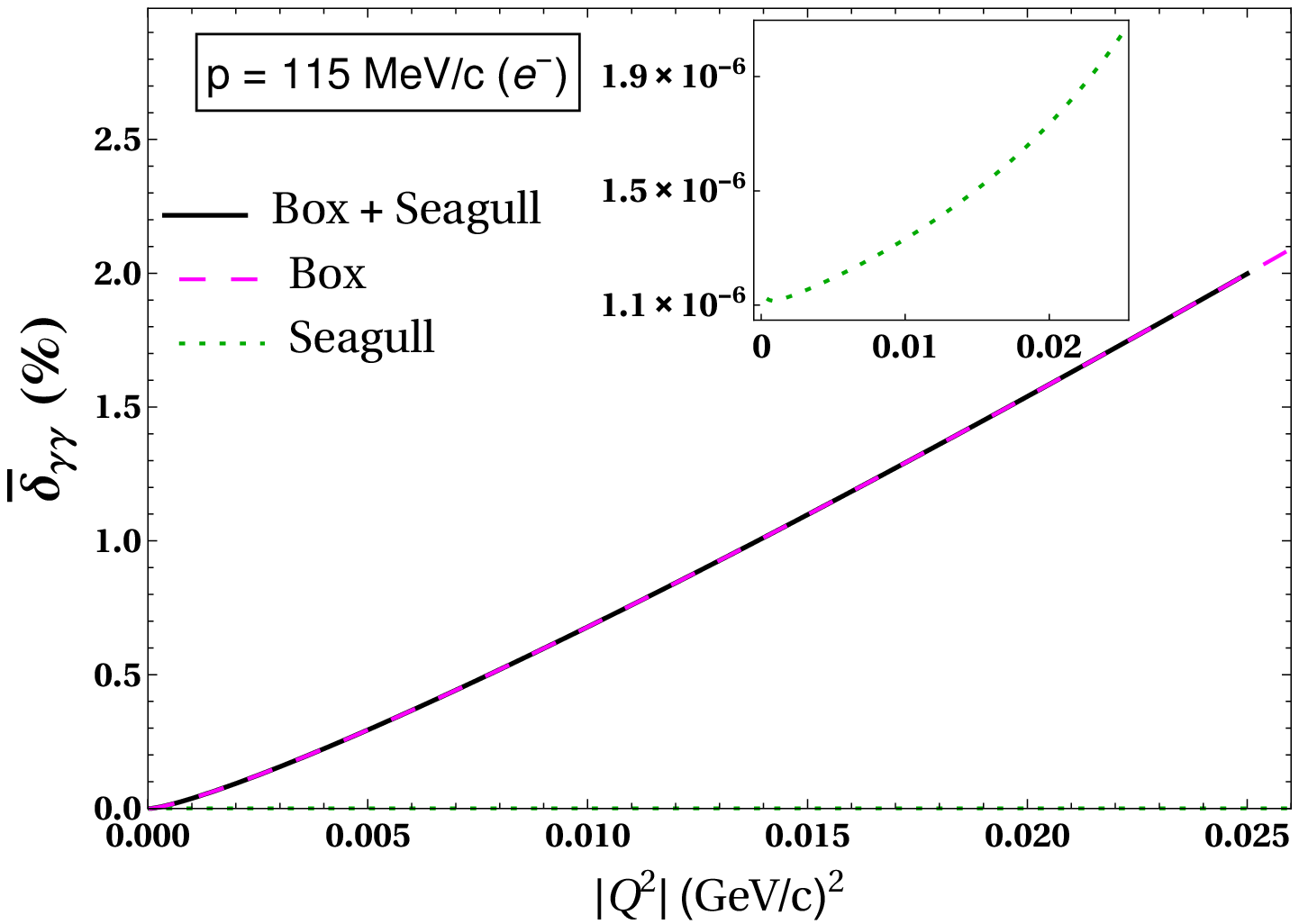}}\qquad\qquad{\includegraphics[scale=0.48]{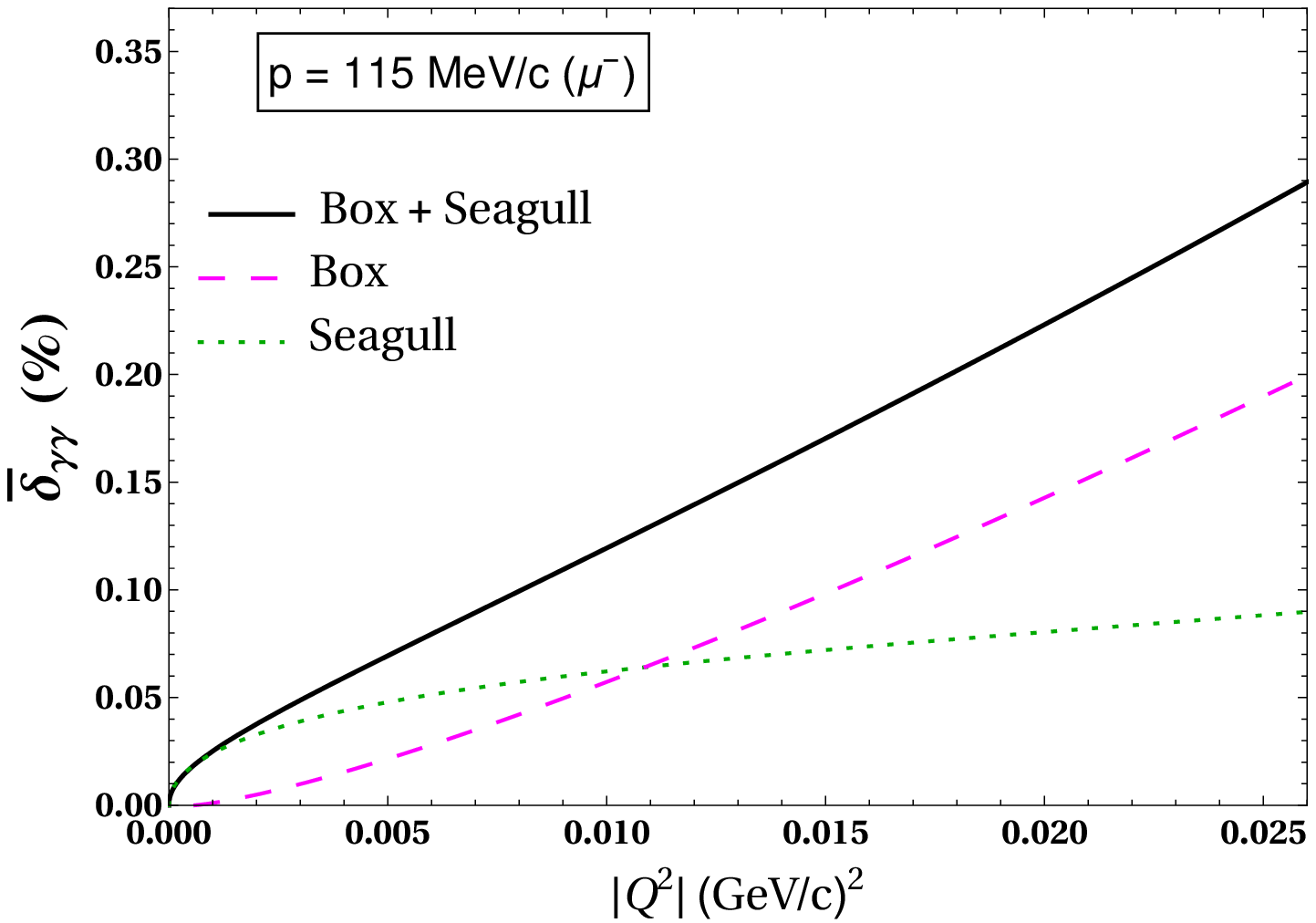}}\\

\vspace{0.4cm}

{\includegraphics[scale=0.48]{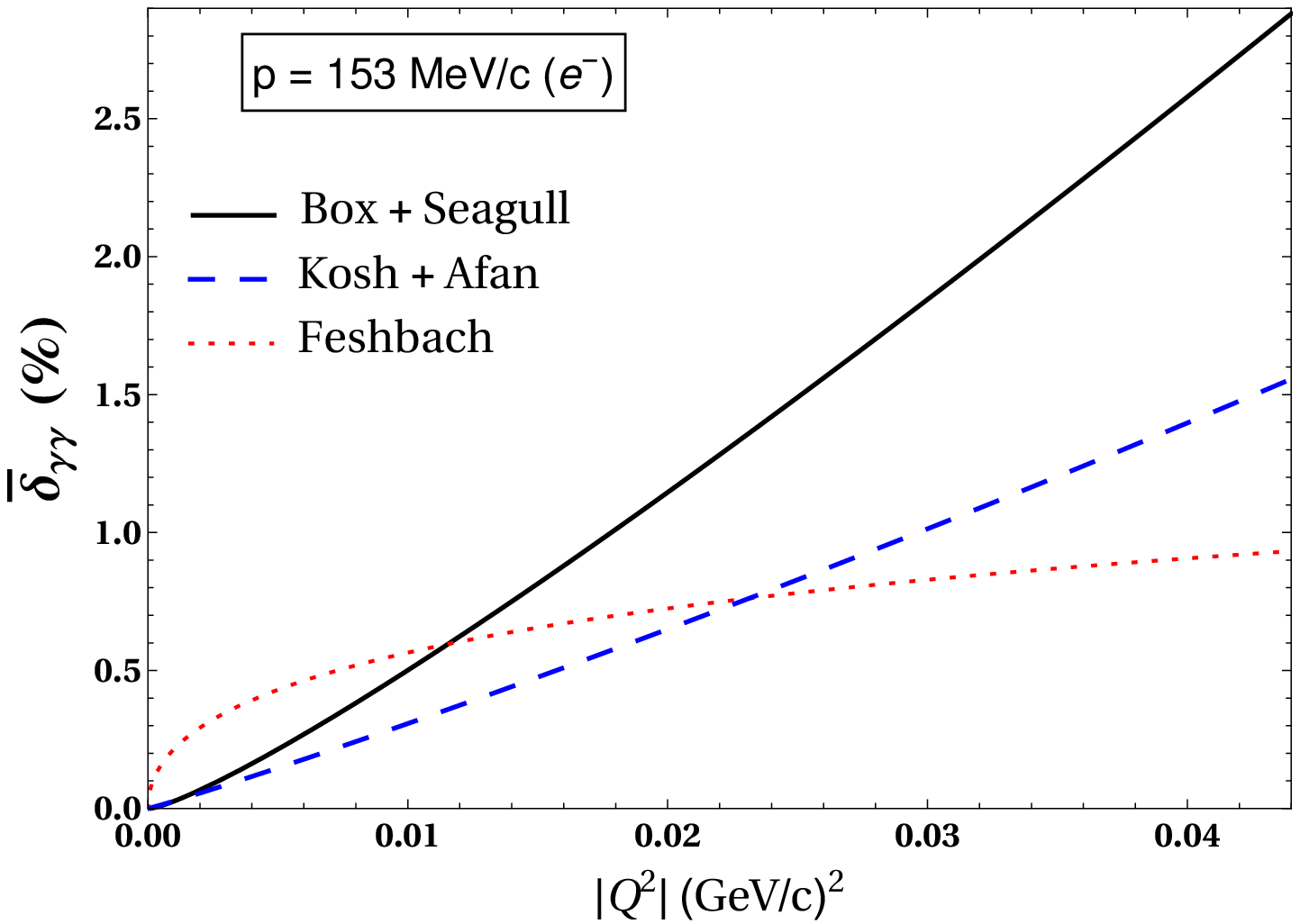}}\qquad\qquad{\includegraphics[scale=0.48]{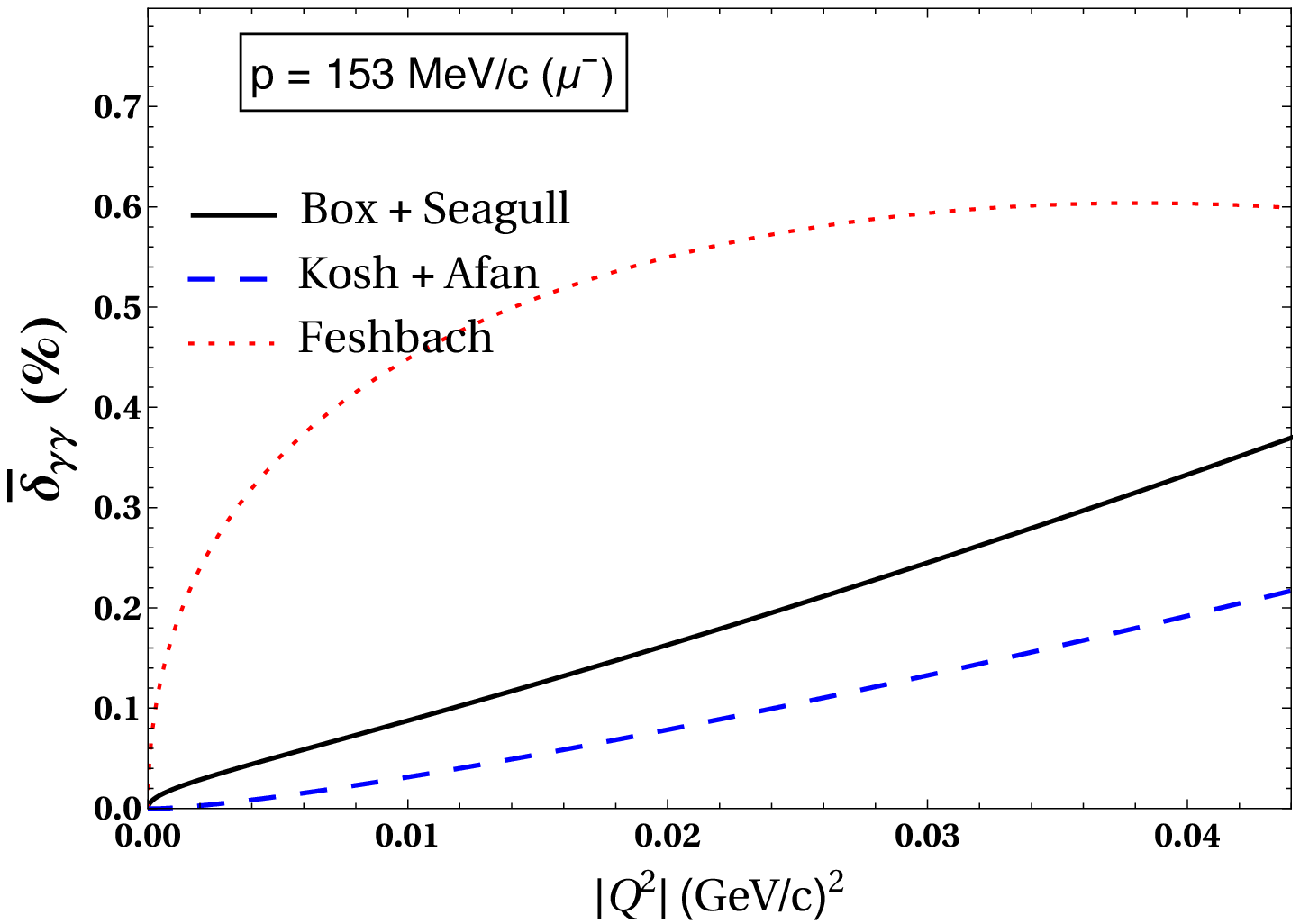}}
\caption{\label{TPE:delta_elmu} Comparison of the finite TPE contributions for the box and seagull 
        diagrams to the $e$p (left panel) and $\mu$p (right panel) elastic scattering cross sections 
        as a function of the squared four-momentum transfer $|Q^2|$ given at the three proposed MUSE 
        incoming lepton momenta, namely, $210$ MeV/c, $153$ MeV/c and $115$ MeV/c. The seagull 
        contributions for the $e$p scattering, being numerically much smaller, are shown within the 
        inset plots. The plots in the bottom panel ($4^{th}$ row) show the comparison of our results 
        (for incoming lepton momentum, $p=153$ MeV/c) with the qualitatively similar TPE results from 
        the recent relativistic hadronic model calculation of Ref.~\cite{Koshchii:2017dzr} (labeled as 
        ``Kosh + Afan''). The contribution of the Feshbach term of Ref.~\cite{McKinley:1948zz} (labeled 
        as ``Feshbach'') is also displayed.}
\end{figure*}

Next we present numerical estimates of the analytically derived expressions for the box and seagull TPE contributions 
obtained in the previous section. Figure.~\ref{TPE:delta_elmu}  displays our results showing the dependence of the 
finite fractional TPE corrections, $\bar{\delta}_{\gamma\gamma}$, of the $e$p and $\mu$p elastic scattering versus the 
squared four-momentum transfer at ${\mathcal O}\left(\alpha,M^{-1}\right)$. The results displayed in the figure indicate 
that the TPE corrections for electron-proton scattering goes up to about $4.5\%$ and that for the muon-proton 
scattering up to around $0.5\,\%$ for the largest MUSE incoming momentum. As anticipated from recent TPE works, e.g.,  
Refs.~\cite{Arrington:2011dn,tomalak2014two,Koshchii:2017dzr}, our TPE contributions are vanishing when $Q^2$ 
becomes zero.\!\!\!~\footnote{It may be noted that in Ref.~\cite{Koshchii:2017dzr} a direct evaluation of the TPE 
(c.f. Eq.~(20) of this reference) leads to a non-zero contribution at $Q^2=0$, and hence, needed to be shifted by a 
constant factor to provide physical justification of vanishing asymmetry at $Q^2=0$. However, we checked that an 
expansion of their Eq.~(20) to ${\cal O}(1/M)$ indeed vanishes at $Q^2=0$.} As observed in the figure, the TPE 
contributions for electron-proton scattering are about an order of magnitude larger than for muon-proton scattering. 
At a given $|Q^2|$ and for increasing incident MUSE lepton momenta, $\bar{\delta}_{\gamma\gamma}$ becomes smaller but the 
relative electron-proton to muon-proton ratio for $\bar{\delta}_{\gamma\gamma}$ stays almost the same. In 
Fig.~\ref{TPE:delta_elmu} we also compare our evaluations of the TPE box, $\bar{\delta}^{\rm (box)}_{\gamma\gamma}$, and the 
seagull, $\delta^{\rm (seagull)}_{\gamma\gamma}$, contributions. Here we note that in relativistic QED the TPE box (and cross-box) 
diagrams give the TPE amplitude, whereas in HB$\chi$PT the baryons being treated non-relativistically, the seagull 
diagram naturally appears. In this case, the magnitude of the finite seagull contribution is found to be quite 
insensitive to the $Q^2$ dependence except when $Q^2 \to 0$. For electron-proton scattering, the seagull contribution 
is more or less inconsequential yielding a minuscule contribution, i.e., $\sim 10^{-6}\,\%$ for the range of MUSE 
kinematics, while for the muon-proton scattering its contribution is much larger going up to about $0.06\,\%$. The TPE 
box diagrams, however, mostly dominate the entire MUSE range of momentum transfers. An exception only occurs for 
muon-proton scattering in the region, $Q^2 \lesssim 0.01$ (GeV/c)$^2$, where our result indicates that the seagull 
terms become numerically larger than the box contributions.

The TPE results of Ref.~\cite{Koshchii:2017dzr}, which we label as ``Kosh + Afan'', are compared with our evaluations 
in the bottom panels of Fig.~\ref{TPE:delta_elmu}, when we adjust their expressions to reflect our IR treatment of the 
TPE amplitude. To be specific, in the method used for comparing the TPE results, we consider only the relevant finite part 
of their TPE result (c.f. Eq.~(20) in Ref.~\cite{Koshchii:2017dzr}), leaving out the IR singular terms which must cancel 
against those from soft bremsstrahlung. In our notation Eq.~(20) of Ref.~\cite{Koshchii:2017dzr} for the TPE (adjusted 
by a constant factor such that it vanishes at $Q^2=0$) in the SPA is written as follows: 
\begin{widetext}
\begin{eqnarray}
\bar{\delta}_{\gamma\gamma}(Q^2)\Big|_{\rm Ref.~[57]}
&=&-\frac{\alpha}{\pi}\left[-\frac{b_{11}}{\gamma_{11}}\left\{\ln\left(\frac{-Q^2}{m_l M}\right)+\frac{1}{2}\ln{\alpha_{12}}\cdot\ln\left(\frac{4\gamma_{11}^2}{m^4_l\alpha_{11}(1-\alpha_{11})^2}\right)\right.\right.
\nonumber\\
&&\left.\left.\hspace{2cm}+\,{\rm Sp}\left(\frac{\alpha_{11}(m^2_l-b_{11}+M^2)}{2\gamma_{11}(1-\alpha_{11})}\right)-{\rm Sp}\left(\frac{m^2_l-b_{11}+M^2}{2\gamma_{11}(1-\alpha_{11})}\right)\right\}\right.
\nonumber\\
&&\left.\hspace{0.8cm}+\,\frac{b_{12}}{\gamma_{12}}\left\{\ln\left(\frac{-Q^2}{m_l M}\right)+\frac{1}{2}\ln{\alpha_{12}}\cdot\ln\left(\frac{4\gamma_{12}^2}
{m^4_l\alpha_{12}(1-\alpha_{12})^2}\right)\right.\right.
\nonumber\\
&&\left.\left.\hspace{2cm}+\,{\rm Sp}\left(\frac{\alpha_{12}(m^2_l-b_{12}+M^2)}{2\gamma_{12}(1-\alpha_{12})}\right)
 -{\rm Sp}\left(\frac{m^2_l-b_{12}+M^2}{2\gamma_{12}(1-\alpha_{12})}\right)\right\}\right]\,,
\label{eq:afanasiev}
\end{eqnarray}  
\end{widetext} 
where 
\begin{eqnarray}
b_{11}=2 E M \quad &,& \quad b_{12}=Q^2+b_{11}\,,
\nonumber\\ 
\alpha_{11}=\frac{b_{11}+2\gamma_{11}}{2m^2_l}\quad &,&\quad \gamma_{11}=\frac{1}{2}\sqrt{b_{11}^2-4m^2_lM^2}\,,
\nonumber\\  
\alpha_{12}=\frac{b_{12}+2\gamma_{12}}{2m^2_l}\quad &,&\quad \gamma_{12}=\frac{1}{2}\sqrt{b_{12}^2-4m^2_lM^2}\,\,.
\end{eqnarray}  
In order to facilitate the comparison with our dimensionally regularized TPE expression $\bar{\delta}_{\gamma\gamma}(Q^2)$ 
[c.f. Eqs.~\eqref{tpe_delta}, \eqref{eq:deltaYY_box} and \eqref{eq:deltaYY_seagull}] we modify their analytically 
regularized IR singular terms proportional to $\ln\lambda^2$, where $\lambda$ is a fictitious photon mass, in the 
following way:   
\begin{eqnarray}
\ln\left(\frac{\lambda^2}{m_l M}\right)\mapsto \ln\left(\frac{-Q^2}{m_l M}\right)+\ln\left(\frac{\lambda^2}{-Q^2}\right)\,.
\end{eqnarray}
Note that the IR-divergent terms proportional to $\ln \lambda^2$ for the TPE correction gets canceled by similar IR 
terms from soft photon bremsstrahlung process leading to their finite expression, Eq.~(38) in Ref.~\cite{Koshchii:2017dzr}.  
We observe in Fig.~\ref{TPE:delta_elmu} that the overall low-$|Q^2|$ behavior of our TPE contributions is roughly 
consistent with Ref.~\cite{Koshchii:2017dzr} SPA results which are based on the use of relativistic point-like (Dirac) 
protons. Nevertheless, despite the apparent qualitative similarity, our total TPE contribution differs in magnitude 
roughly by about a factor of two from that in Ref.~\cite{Koshchii:2017dzr}. Moreover, we note that our results 
substantially differ from the results of another recent TPE work, Ref.~\cite{tomalak2014two}, which evaluated the box 
diagrams for muon-proton scattering using a relativistic hadronic model. However, unlike Ref.~\cite{Koshchii:2017dzr} 
and our work, the authors of Ref.~\cite{tomalak2014two} did not employ SPA in their calculations, and instead 
numerically evaluated the TPE amplitudes involving the so-called 
{\it four-point integrals}~\cite{hooft:1979,passarino:1979} and their derivatives. In addition, they isolated the IR 
singular terms analytically from their TPE amplitude according to the Maximon and Tjon prescription~\cite{Maximon:2000hm}. 
The significant difference of our TPE correction as well as the results of Ref.~\cite{Koshchii:2017dzr} from those in 
Ref.~\cite{tomalak2014two} may imply that a part of the TPE box diagram loop integration involves contributions from 
two ``hard'' photon exchanges in muon-proton scattering. This is precisely the integration region of these TPE loops 
excluded in SPA. 

Furthermore, in Fig.~\ref{TPE:delta_elmu} we compare our TPE results with the Coulomb potential scattering result in 
the second Born approximation by McKinley and Feshbach~\cite{McKinley:1948zz}, labeled ``Feshbach'' in the figure. 
As shown in Refs.~\cite{tomalak2014two,Tomalak:2015hva}, the relativistic evaluation of the TPE diagrams for a 
point-like Dirac proton without SPA are qualitatively very similar to the Feshbach contribution for muon-proton scattering.  
Nevertheless, it may be worth noting that the original Feshbach derivation is applicable only for ultra-relativistic 
electrons. As seen in Fig.~\ref{TPE:delta_elmu} for the electron-proton scattering, our results as well as those in 
Ref.~\cite{Koshchii:2017dzr} are comparable to the Feshbach term for low-$|Q^2|$ values, thereby indicating that the 
``hard'' photon TPE loop contributions might not be too important for electron-proton scattering.  

\begin{figure}[tbp]
\centering
{\includegraphics[scale=0.48]{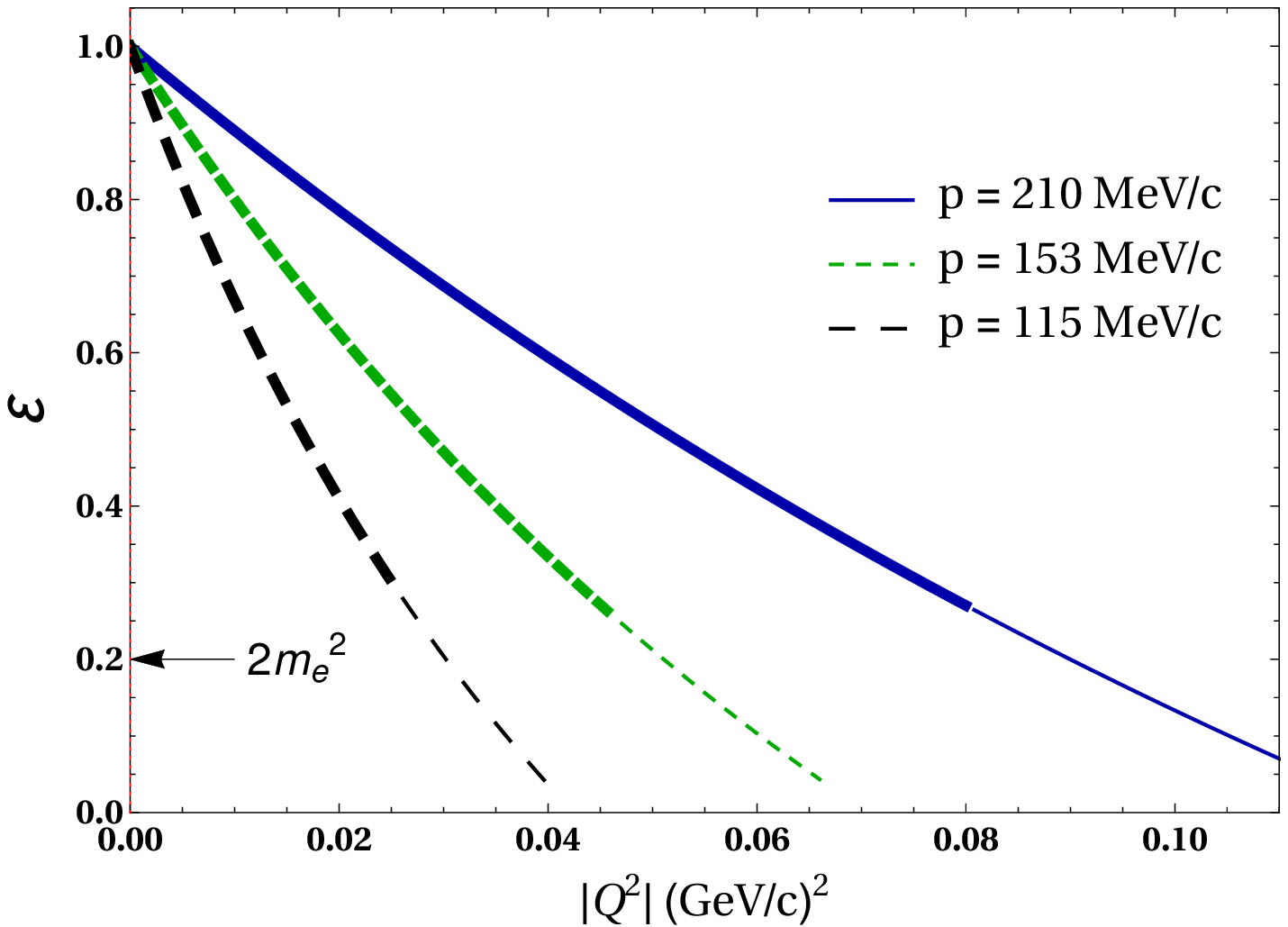}}

\vspace{0.4cm}

{\includegraphics[scale=0.48]{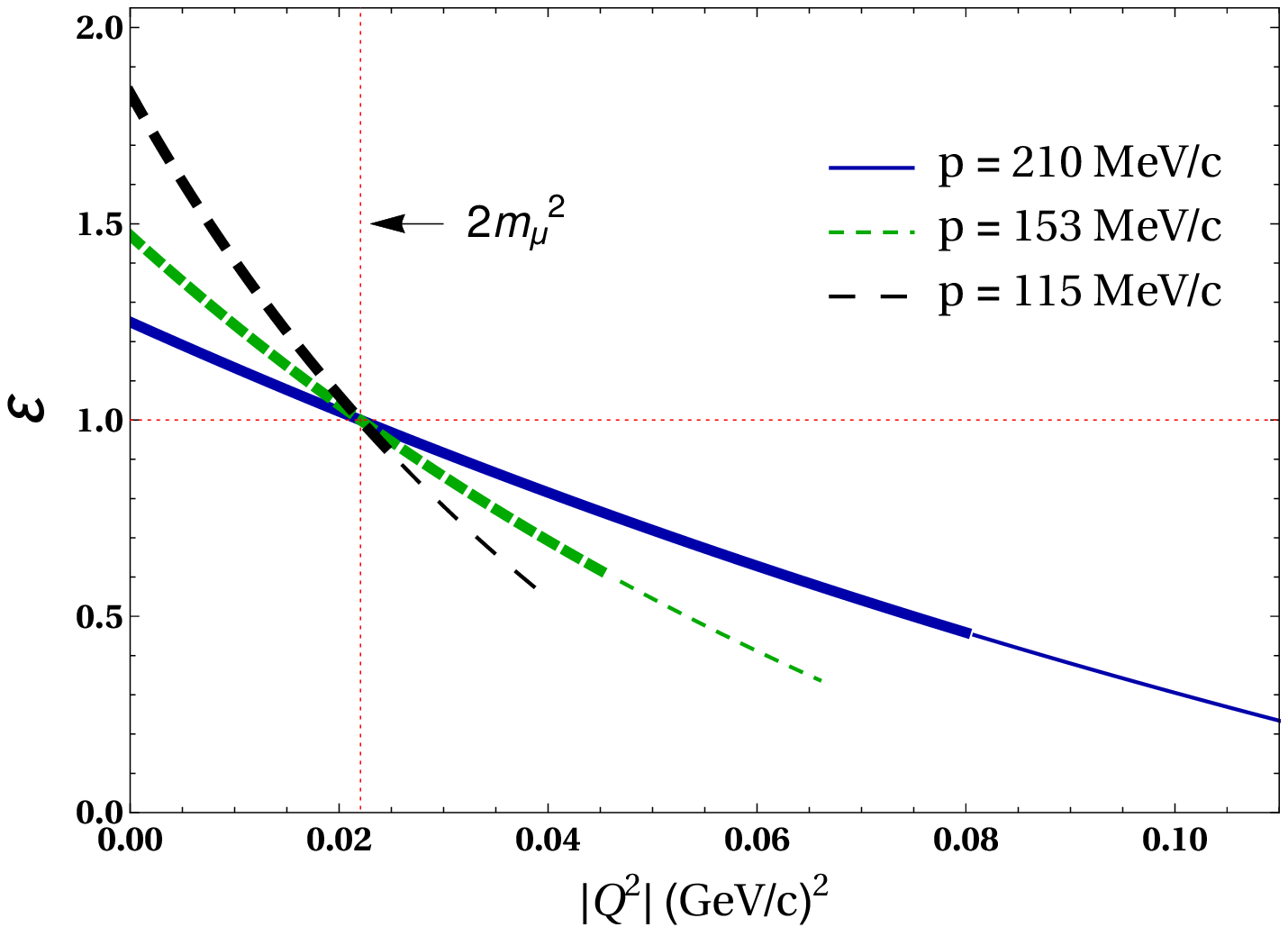}}
\caption{\label{epsilon_Q2} The dependence of the virtual photon ``polarization'' factor $\varepsilon$ 
        on the  squared four-momentum transfer $|Q^2|$ for the proposed MUSE beam momenta 
        for $e$p scattering (upper panel) and $\mu$p scattering (lower panel.) Each plot corresponds 
        to the full kinematically allowed scattering range $0<|Q^2|< |Q^2_{\rm max}|$ when 
        $\theta\in [0,\pi]$ (thin lines). The thick lines are associated with the MUSE kinematic 
        range  where $\theta\in [20^{\circ},100^{\circ}]$. The curves intersect at $\varepsilon=1$,  which 
        correspond to the critical values, $|Q^2_{\rm crit}|=2m^2_e=5\times 10^{-7}$ (GeV/c)$^2$ and 
        $|Q^2_{\rm crit}|=2m^2_{\mu}=0.02205$ (GeV/c)$^2$.}
\end{figure}

\begin{figure*}[tbp]
\centering
{\includegraphics[scale=0.48]{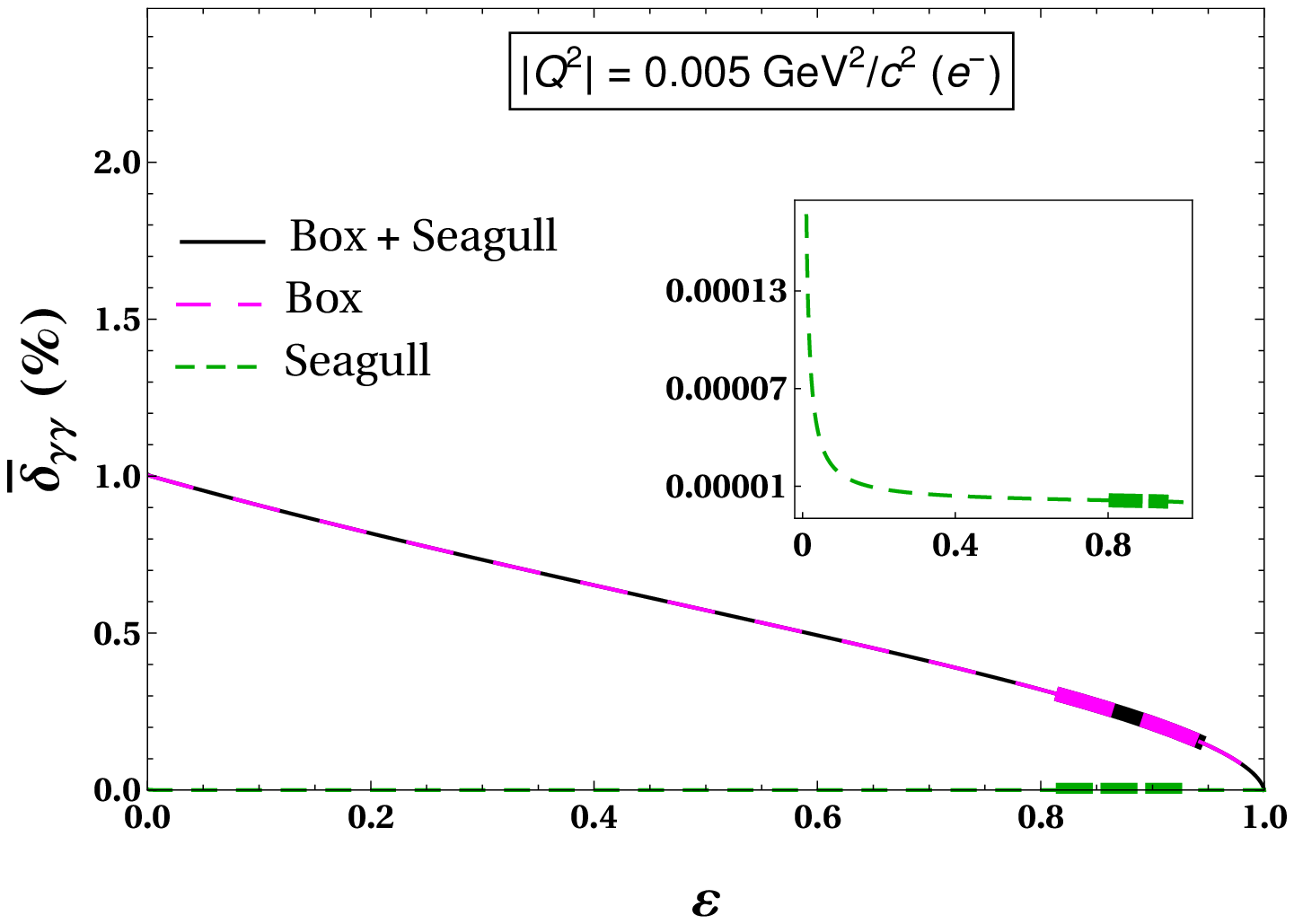}}\qquad\qquad{\includegraphics[scale=0.48]{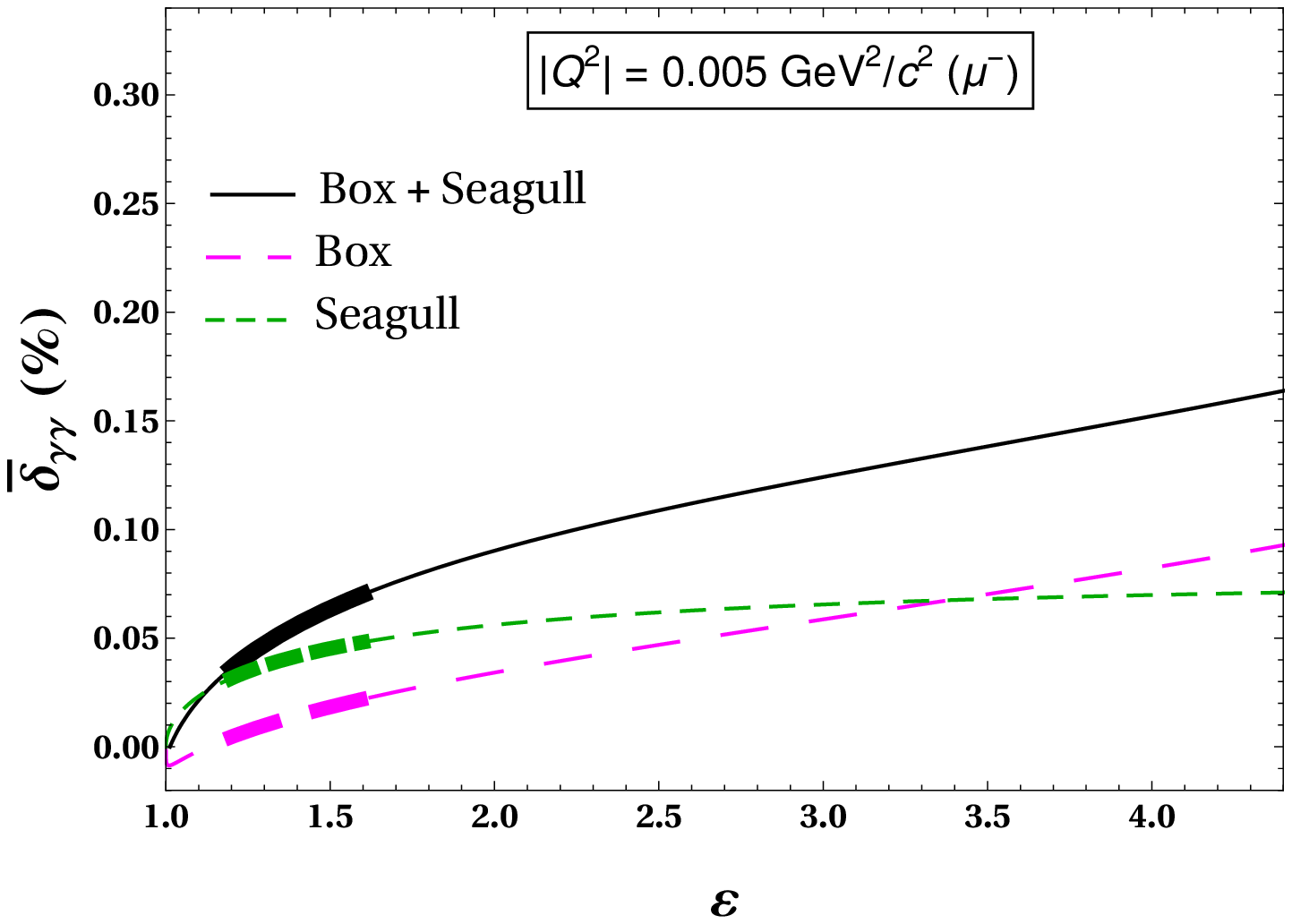}}\\

\vspace{0.4cm}

{\includegraphics[scale=0.48]{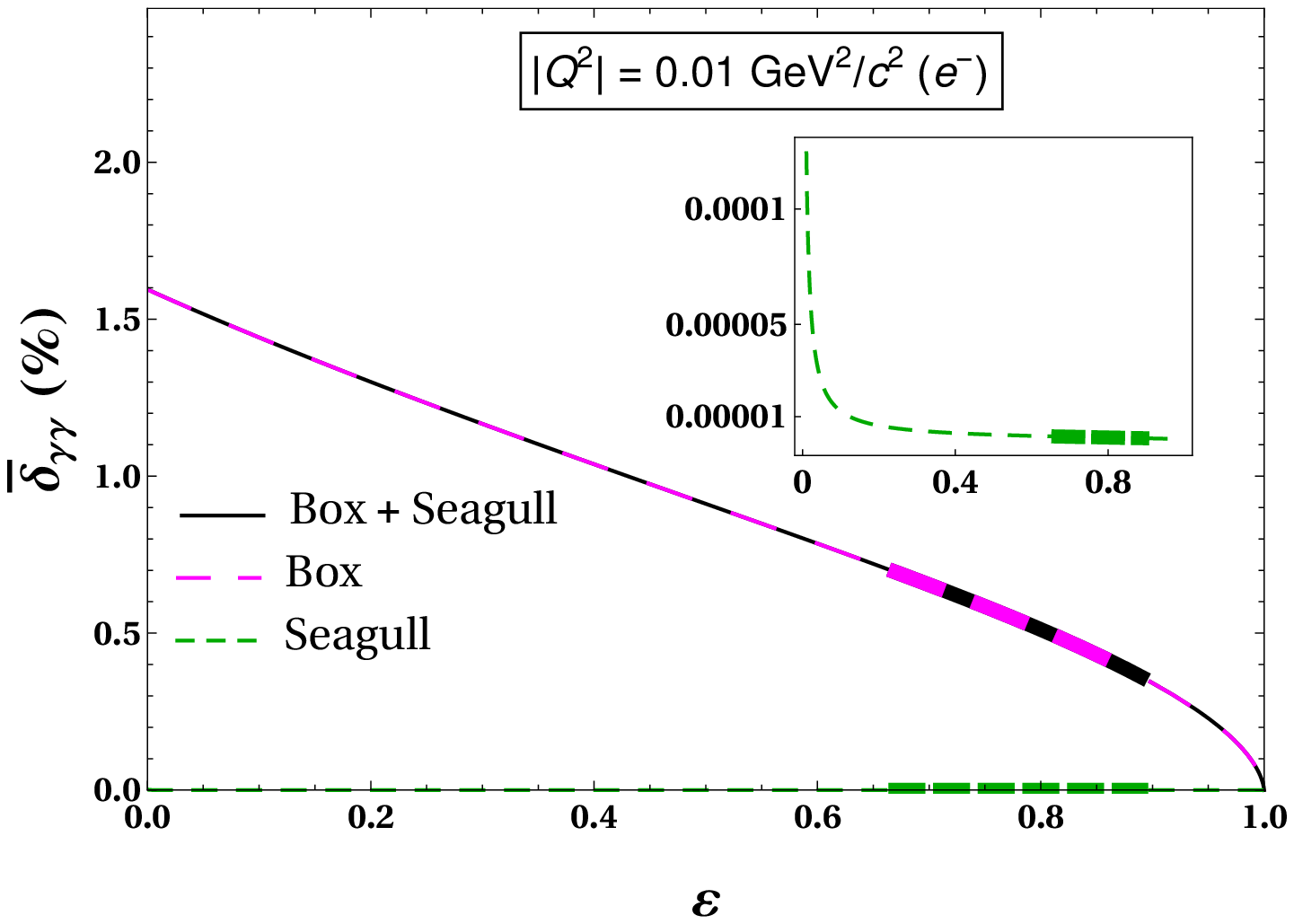}}\qquad\qquad{\includegraphics[scale=0.48]{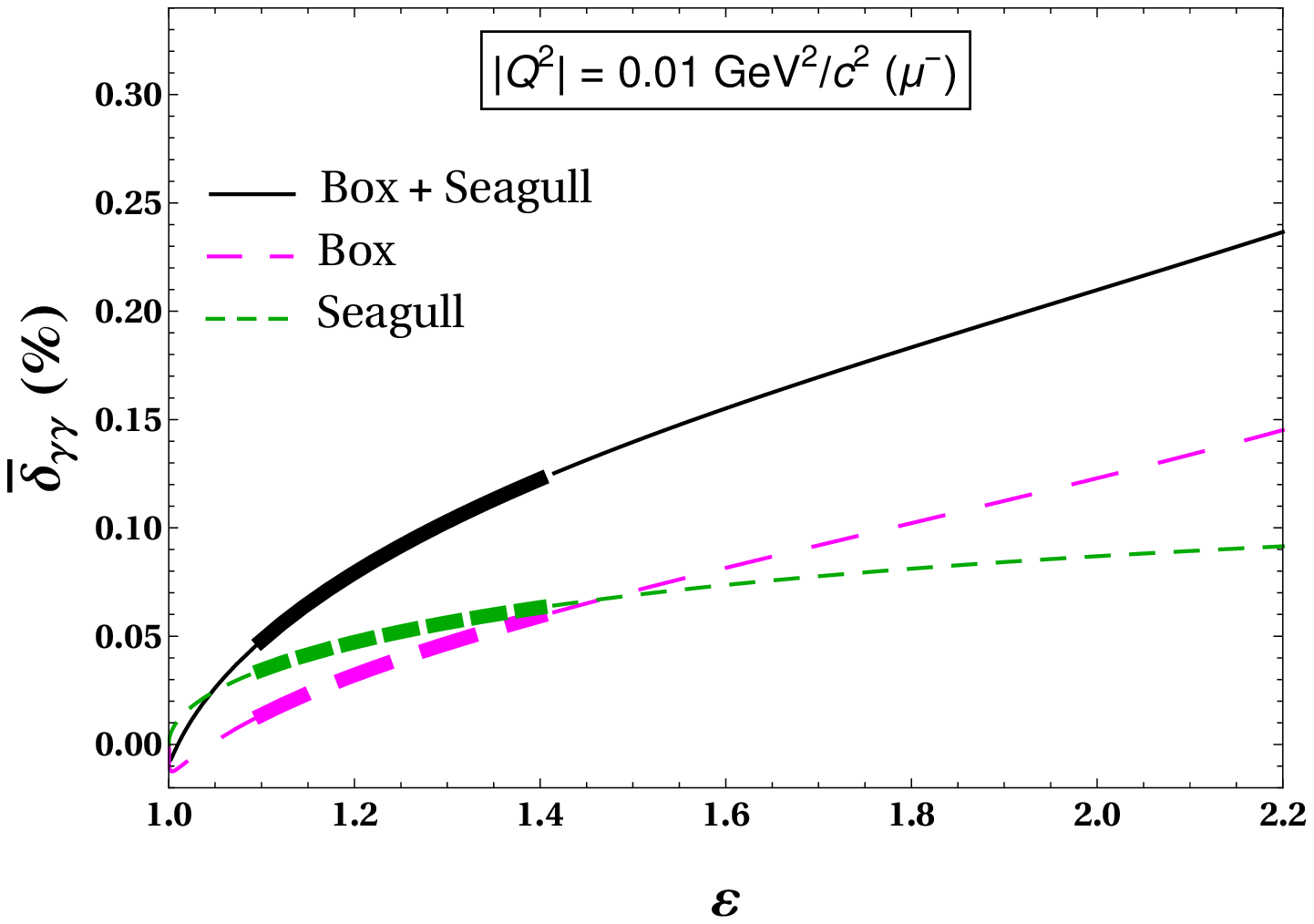}}\\

\vspace{0.4cm}

{\includegraphics[scale=0.48]{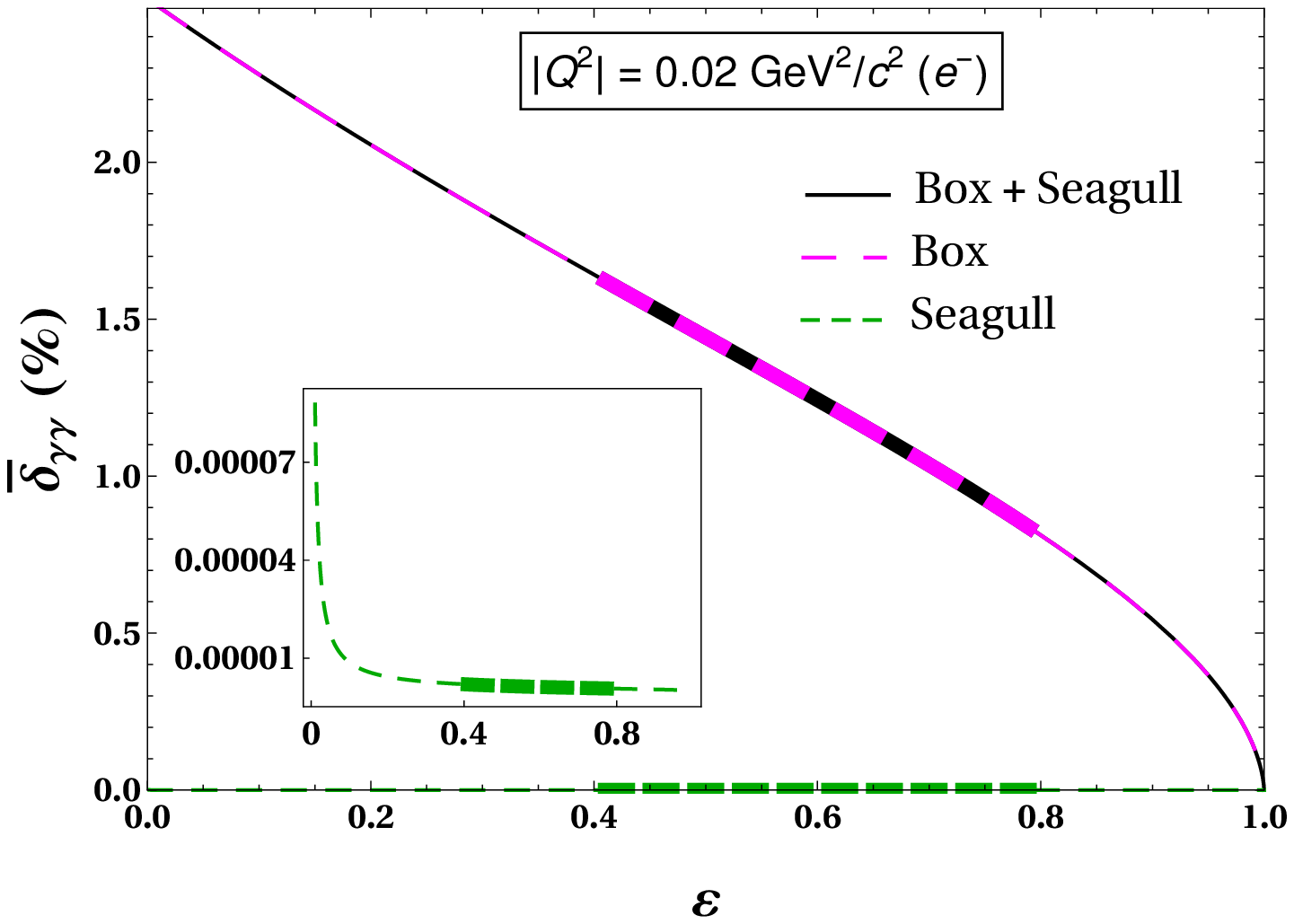}}\qquad\qquad{\includegraphics[scale=0.48]{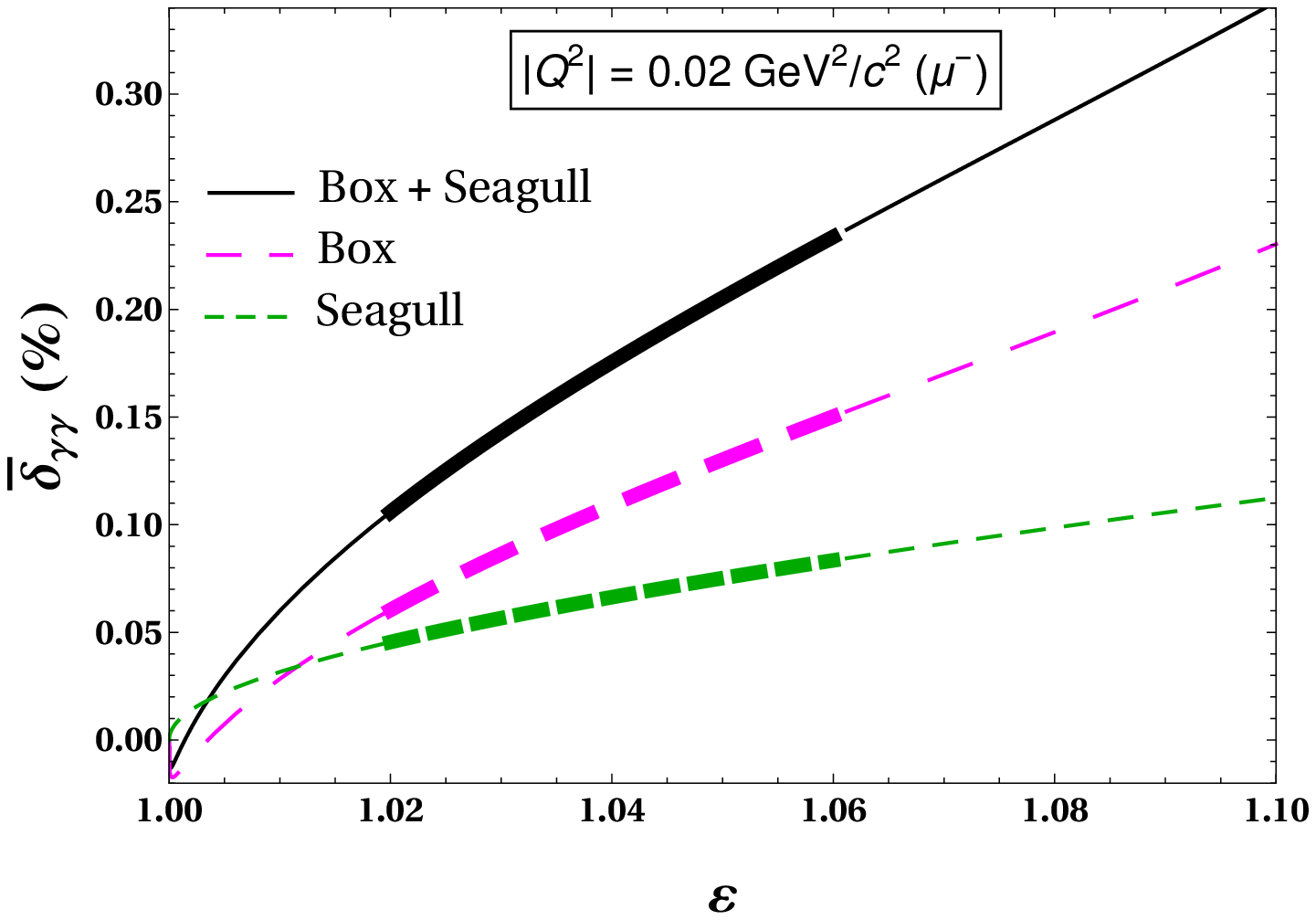}}\\

\vspace{0.4cm}

{\includegraphics[scale=0.48]{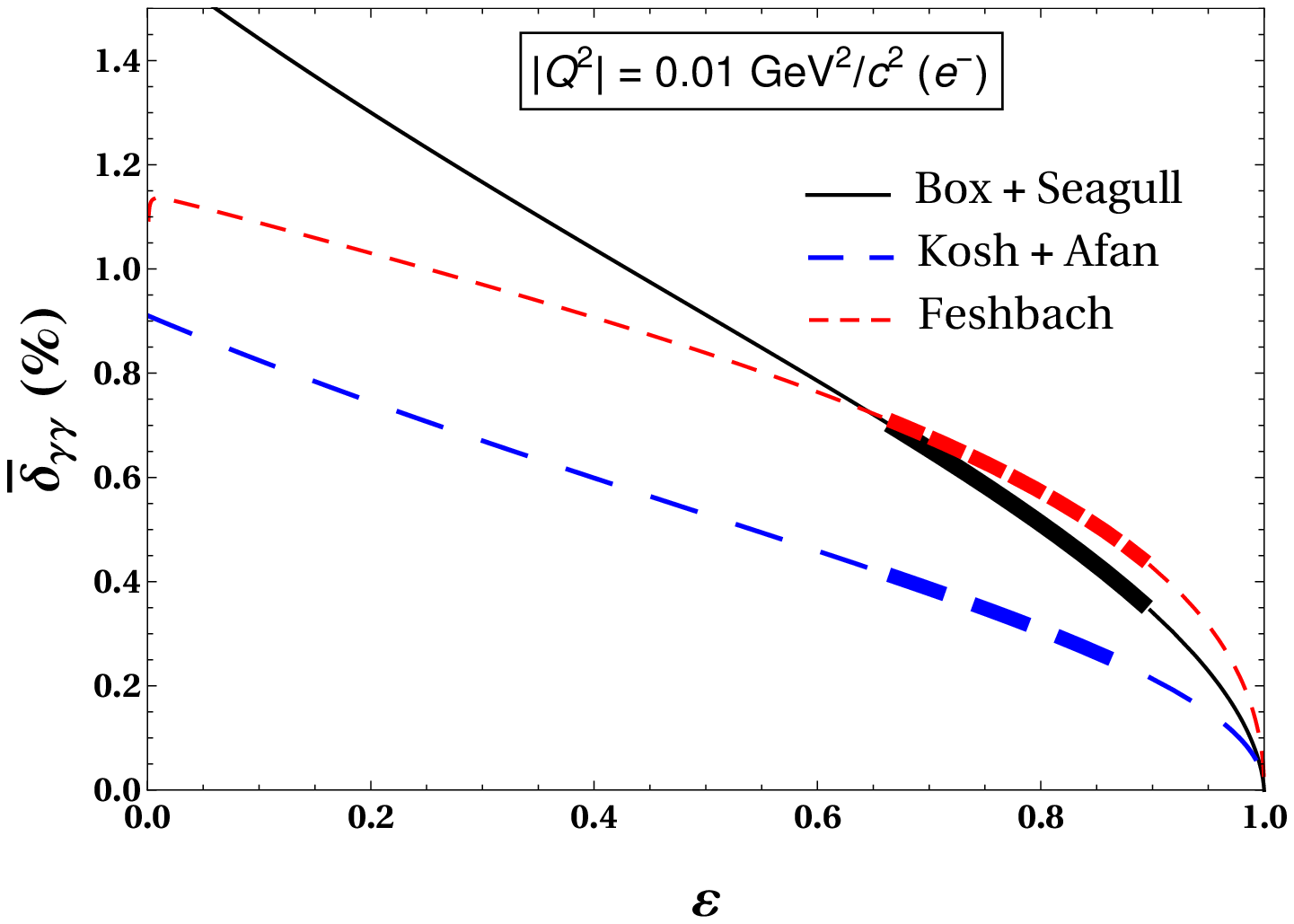}}\qquad\qquad{\includegraphics[scale=0.48]{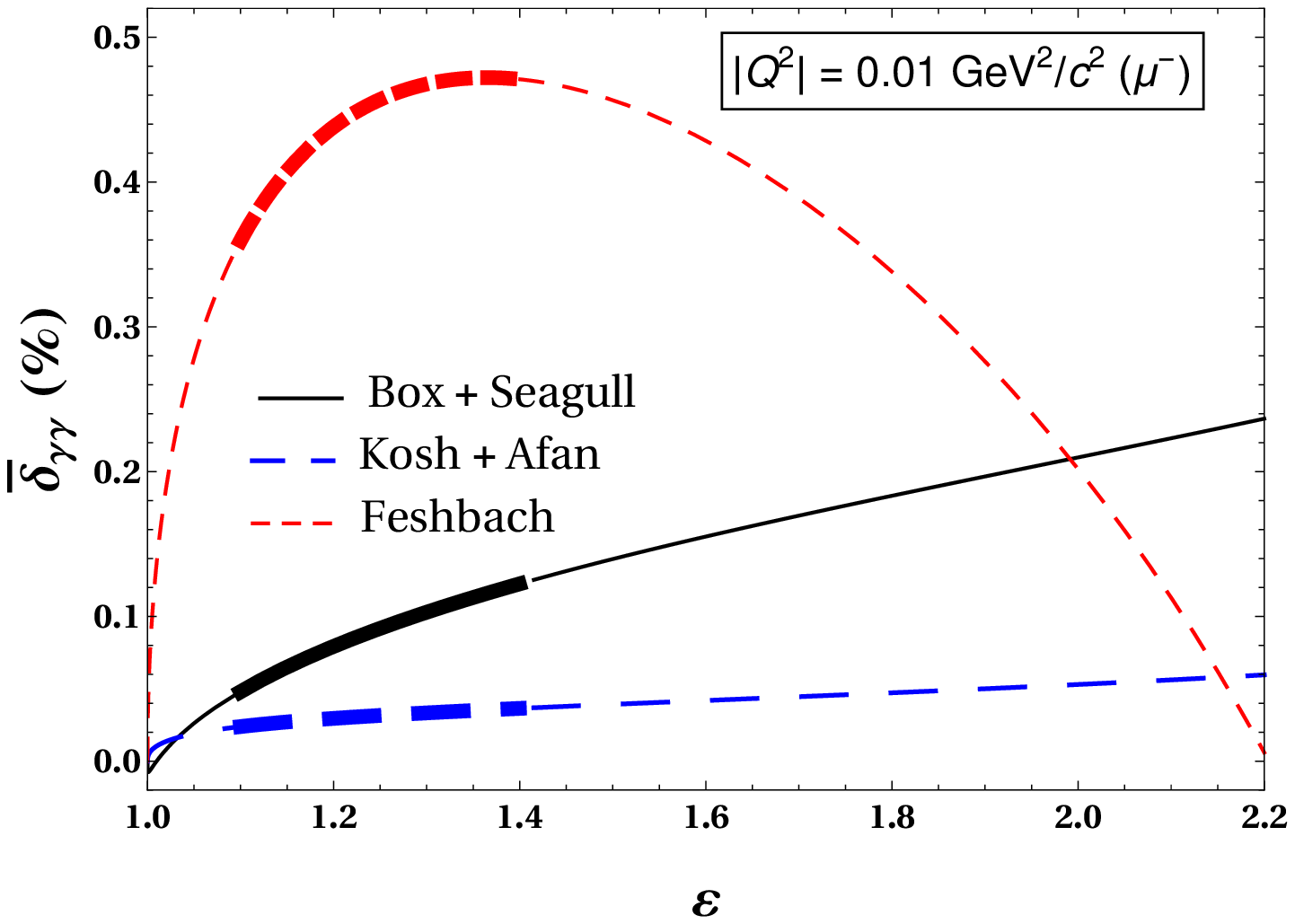}}
\caption{\label{TPE:delta_elmu_epsilon} The $\varepsilon$ dependence of the finite TPE contributions 
        for $e$p (left panel) and $\mu$p (right panel) elastic cross sections for three specific $|Q^2|$ 
        values in the proposed MUSE kinematic range. The seagull contributions for the $e$p scattering 
        being numerically much smaller are shown within the inset plots. The plots in the bottom panel 
        ($4^{th}$ row) show the comparison of our results for $|Q^2|=0.01$ (GeV/c)$^2$ with the 
        qualitatively similar results from Ref.~\cite{Koshchii:2017dzr} (labeled as ``Kosh + Afan''.) 
        The contribution of the Feshbach term~\cite{McKinley:1948zz} (labeled as ``Feshbach'') is also 
        displayed. Each plot corresponds to the kinematically allowed range of $\varepsilon$ when 
        $\theta\in [0,\pi]$ (thin lines), and the `segment' relevant to the MUSE kinematic range with 
        $\theta\in [20^{\circ},100^{\circ}]$ (thick lines).} 
\end{figure*}

It is also instructive to study the TPE dependence on the virtual photon ``polarization'' flux factor 
$\varepsilon$ which may be expressed in terms of the four-momentum transfer $Q^2$ by the relation~\cite{tomalak2014two}
\begin{equation}
\varepsilon(Q^2) =\frac{16\nu^2+Q^2(4M^2-Q^2)}{16\nu^2-(4M^2-Q^2)(4m_l^2+Q^2)}\,,
\end{equation}
where $\nu=(s-u)/4=(4EM+Q^2)/4$ is the crossing symmetric variable in the target rest frame. For fixed incident 
lepton beam momenta, the full kinematically allowed elastic scattering range, namely, $0<\theta<\pi$ and 
$0<|Q^2|< |Q^2_{\rm max}|$ [c.f. below Eq.~\eqref{eq:Q^2}], yields the physical bound on the flux factor, 
namely, $\varepsilon_{\rm max} >\varepsilon >\varepsilon_{\rm min}$, where
\begin{eqnarray}
\varepsilon_{\rm max}&\equiv&\varepsilon(0)=\frac{1}{\beta^2}\,,
\nonumber\\
\varepsilon_{\rm min}&\equiv&\varepsilon(Q^2_{\rm max})=\frac{m^2_l(m^2_l+M^2+2EM)}{2\beta^2E^2M^2}\,.
\end{eqnarray}
While for fixed four-momentum transfers, if $|Q^2|>2m^2_l$, then $2m^2_l/|Q^2|<\varepsilon<1$, and if $|Q^2|<2m^2_l$, 
then $1<\varepsilon<2m^2_l/|Q^2|$. The critical case, $|Q^2|=|Q^2_{\rm crit}|=2m^2_l$ corresponds to $\varepsilon=1$ 
for all possible incoming lepton momenta. It is worth noting that for the massless lepton case, $\varepsilon$ may be 
interpreted as the longitudinal polarization of the photon in case of one-photon exchange~\cite{tomalak2014two}. 
Figure.~\ref{epsilon_Q2} displays the $|Q^2|$ dependence of $\varepsilon$ for $e$p and $\mu$p elastic scatterings. The 
figure identifies both the kinematically allowed and the relevant MUSE range of $\varepsilon$ values. Correspondingly, 
Fig.~\ref{TPE:delta_elmu_epsilon} displays the $\varepsilon$ dependence of our TPE corrections for three specific 
choices of $|Q^2|$, namely, $0.005$ (GeV/c)$^2$, $0.01$ (GeV/c)$^2$ and $0.02$ (GeV/c)$^2$. 
In each case of fixed $|Q^2|$ the TPE effects vanish as $\varepsilon\to 1$, i.e., the forward scattering limit, and tend 
toward maximum as $\varepsilon\to 2m^2_l/|Q^2|$ for backward scatterings~\cite{Arrington:2011dn}, as reflected in 
Fig.~\ref{TPE:delta_elmu_epsilon}. This feature of our TPE result is again qualitatively similar to the result obtained 
in Ref.~\cite{Koshchii:2017dzr} but contrasts sharply with the Feshbach result~\cite{McKinley:1948zz} as well as that of 
Ref.~\cite{tomalak2014two}. 

\section{Summary and Conclusion}
\label{conclusion}
We present a low-energy model-independent calculation of the two-photon exchange contributions 
to the lepton-proton elastic unpolarized cross section at next-to-leading order in HB$\chi$PT. 
The lepton mass is included in all our expressions. Our approach contrasts many previous TPE evaluations 
using relativistic hadronic models which often use phenomenological form factors to parametrize the 
proton-photon vertices. In HB$\chi$PT the heavy proton is treated in a manifestly non-relativistic 
framework which makes it ideal for investigating the structure of the proton at very low momentum transfers. 
Our evaluation is based on the assumption that the most dominant contributions to the TPE loop diagrams 
arise from the elastic proton intermediate state while inelastic contributions are considered small for 
low-$|Q^2|$ values. This is especially relevant in the proposed low-energy MUSE kinematic domain where 
incoming lepton beam momenta are between $p=115 and 210$ MeV/c. We note that while most other works use 
analytic regularization schemes with a non-zero photon mass, we used the gauge invariant prescription of 
dimensional regularization scheme to isolate the infrared singularities of the two-photon loops. In this 
approach, however, the exact evaluation of the IR-divergent four-point one-loop Green's 
function~\cite{hooft:1979,passarino:1979}  demands analytical evaluations of $D$-dimensional integrals 
which to the best of our knowledge have not been pursued for exchanges of massless photons. Further, we 
demonstrated in Sec.~\ref{TPE}  that the soft photon limit was taken only after the cancellations among 
the NLO amplitudes were taken into account. Moreover, one should bear in mind that we restricted 
the soft photon approximation only to the IR-divergent diagrams. Thus, we evaluated the IR-free seagull 
diagram without invoking the soft photon approximation. 

The evaluation of the TPE box diagrams (with exchange bosons with non-zero masses) involves scalar and 
tensor four-point loop integrals, a topic that has been discussed extensively in many previous works, e.g., 
Refs.~\cite{hooft:1979,passarino:1979,beneke:1998,grozin:2001,zupan:2002}. In the pioneering works of 
Refs.~\cite{hooft:1979,passarino:1979}, such tensor integrals were reduced to scalar one-loop {\it master} 
integrals involving one-, two-, three- and four-point functions which were evaluated analytically using 
the dimensional regularization scheme. The work of Ref.~\cite{zupan:2002} extended the above formalism to 
include heavy-fermion propagators. However, these approaches are unsuitable for dealing with IR divergences 
using dimensional regularization with massless photon exchanges. In fact, for massless photon exchanges 
the exact analytical evaluation of the IR-divergent four-point functions in dimensional regularization 
remains an open issue. Nevertheless, using dimensional regularization with the soft photon approximation 
provides a viable alternative for the reduction of the four-point scalar integrals into well-known standard 
ones. This approximation allows us to easily project out the IR-divergent parts of the TPE box diagrams in 
order to obtain a finite contribution to the elastic cross section. In the soft photon limit the four-point 
loop integral reduces to a three-point integrals which can be evaluated analytically, wherein each of the 
TPE loop momentum, $0 \leq (k_0,|{\bf k}|) \leq \infty$, can be decomposed as a sum of two integrals each 
with a hard and a soft photon. The contributions from two simultaneous hard photon exchanges are ignored in 
the soft photon limit. 

The results for the electron-proton scattering seem to indicate that the dominant contribution from the TPE 
loop momenta are expected to arise from the integration domain where the contribution of the two hard photon 
exchanges may give small contributions. In contrast, it appears that for muon-proton scattering the hard 
two-photon exchange part of the TPE loops could give significant contributions. This conclusion is based on 
a comparison with the muon-proton scattering analysis presented in Ref.~\cite{Tomalak:2015hva} (c.f. Fig. 2 
of this reference), where the TPE amplitude was evaluated relativistically with point-like protons. This may 
indicate the importance of including two hard photon exchanges even in very low-energy muon-proton 
scattering for a more robust estimation of the TPE contribution. 

In conclusion, we demonstrated in this paper many cancellations among the NLO box and seagull diagrams, 
which are likely to remain approximately valid beyond SPA. Furthermore, we showed that while the LO TPE 
contributions vanish (up to an irrelevant imaginary part), the dominant TPE effects arise from the box 
diagrams with NLO proton-photon vertices except at very low-$|Q^2|$ values where the finite seagull terms 
become significantly large for muon-proton scattering. However, the seagull diagram gives only a tiny 
contribution for electron-proton scattering at MUSE energies. We find that the low-$|Q^2|$ behavior of 
our TPE contributions are in rough agreement with the results in Ref.~\cite{Koshchii:2017dzr}, although 
they differ substantially from those in Ref.~\cite{tomalak2014two}. As a next step it could be desirable 
to use a standard numerical packages in order to evaluate the TPE box diagrams in the heavy baryon scheme 
and to examine the robustness of our soft photon approximation results. An estimate of the NNLO corrections 
in HB$\chi$PT that include, e.g., contribution of pion loops and the proton's magnetic moment, would also 
be helpful in understanding the uncertainties using the HB$\chi$PT approach. Additionally, the inelastic 
$\Delta$ intermediate TPE contributions may be included to constrain uncertainties.   

\vspace{-0.6cm}

\section*{Acknowledgments}
The authors thank Steffen Strauch for various useful discussions relating 
to the the MUSE Collaboration. P.T. thanks the Department of Physics and 
Astronomy, University of South Carolina, for kind hospitality during the 
initial stages of this work. V.S. thanks the Department of Physics, Indian 
Institute of Technology Guwahati, for kind hospitality at various stages 
of this work. 

\vspace{-0.4cm}

\section*{Appendix}
Here we derive the TPE seagull amplitude ${\mathcal M}^{(i)}_{\rm seagull}$ from  diagram (i) in Fig.~\ref{tpe}.
The integral in Eq.~\eqref{M_i} being both IR and UV finite does not require any regularization and is   
evaluated analytically in the 4-dimensional space-time. In section \ref{TPE}, it was shown {\it without} 
implementing SPA that due to cancellations among various NLO TPE amplitudes only the residual part of 
$\widetilde{\mathcal M}^{(i)}_{\rm seagull}$ proportional to $g^{\mu\nu}$ contributes to the cross section, 
namely,
\begin{eqnarray}
\widetilde{\mathcal M}^{(i)}_{\rm seagull}&=&-\,\frac{e^4}{M}\frac{1}{i}\int \frac{d^4k}{(2\pi)^4}
\nonumber\\
&\times& \frac{\left[\bar u(p^\prime)\gamma^\mu(\slashed{p}-\slashed{k}+m_l)\gamma_\mu u(p)\right]\,
\left[\chi^\dagger(p_p^\prime)\chi(p_p)\right]}{k^2\,(Q-k)^2\,(k^2-2k\cdot p+i0)}
\nonumber\\
&=&-\,\frac{e^4}{M}[\bar u(p^\prime)\gamma^\alpha\, {\mathbb I}_{\rm \,seagull}\, \gamma_\alpha u(p)]\,\,[\chi^\dagger(p_p^\prime)\chi(p_p)]\,,
\nonumber\\
\label{eq:seagull_amplitude}
 \end{eqnarray}  
where the loop-integral appearing above is given by \\

\vspace{-0.6cm}

\begin{eqnarray}
{\mathbb I}_{\rm \,seagull}=\frac{1}{i}\int \frac{d^4k}{(2\pi)^4}\frac{\slashed{p}-\slashed{k}+m_l}{(k^2+i0)(Q-k)^2(k^2-2k\cdot p+i0)}
\nonumber
\end{eqnarray}\\
where $k$ and $Q-k$ are the momenta of the two exchanged photons and $Q=p-p^\prime$. Using Feynman parametrization 
and thereby shifting the integration variable, $k\to k+\beta$ where $\beta=(px+Qy)$ yields,\\
\begin{eqnarray}
{\mathbb I}_{\rm \,seagull}&=&\frac{1}{i}\int_0^1\int_0^1\int_0^1 dx\, dy\, dz\, \delta(1-x-y-z)
\nonumber\\
&&\hspace{1.5cm} \times \int\frac{d^4k}{(2\pi)^4}\frac{2(\slashed{p}-\slashed{k}-\slashed{\beta}+m_l)}{[k^2-\Delta]^3}\,.
\nonumber
\end{eqnarray}
Integration over the loop momentum $k$, we obtain
\begin{eqnarray}
{\mathbb I}_{\rm \,seagull}&=&-\frac{1}{16\pi^2}\int_0^1\int_0^1\int_0^1 dx\, dy\, dz\, \delta(1-x-y-z) 
\nonumber\\
&&\hspace{1.5cm} \times \left[\frac{\slashed{p}-\slashed{\beta}+m_l}{\Delta}\right] \,,
\end{eqnarray}
where $\Delta=(px+Qy)^2-Q^2y$. Further, it is convenient to make a change of variables from $(x,y,z)$ 
to $(\omega,\xi)$ using the transformation  $x=\omega\xi$, $y=\omega(1-\xi)$ and $z=1-\omega$, which 
amounts to the following change of the integration measures, namely, 
$dx\,dy\,dz\,\delta(1-x-y-z) \mapsto \omega\, d\omega\, d\xi$. 
Thus, we obtain
\begin{widetext} 
\begin{eqnarray}
{\mathbb I}_{\rm \,seagull}&=&-\frac{1}{16\pi^2}\int_0^1\int_0^1d\omega d\xi\, \frac{m_l+\slashed{p}(1-\omega\xi)-\slashed{Q}\omega(1-\xi)}{[\omega(m^2_l\xi^2-Q^2\xi+Q^2)-Q^2(1-\xi)]}
\nonumber\\
\nonumber\\
&=&-\frac{1}{16\pi^2} \int_0^1\int_0^1d\omega d\xi\, \Big[\frac{m_l+\slashed{p}}{\omega (m^2_l\xi^2-Q^2\xi+Q^2)+Q^2(\xi-1)}-\frac{\omega[\slashed{Q}+\xi(\slashed{p}-\slashed{Q})]}{\omega (m^2_l\xi^2-Q^2\xi+Q^2)+Q^2(\xi-1)}\Big]
\nonumber\\
\nonumber\\
&=&-\frac{1}{16\pi^2m^2_l}\left[(m_l+\slashed{p}){\mathcal I}_1-(\slashed{p}-\slashed{Q})\left({\mathcal I}_2+\frac{Q^2}{m^2_l}{\mathcal I}_3\right)-(2\slashed{Q}-\slashed{p})\frac{Q^2}{m^2_l}{\mathcal I}_4+\slashed{Q}\left(\frac{Q^2}{m^2_l}{\mathcal I}_5-{\mathcal I}_6\right)\right]
\end{eqnarray}
\end{widetext} 
with
\begin{eqnarray}
{\mathcal I}_1(Q^2)&=&\int^1_0 {\rm d}\xi\left[\frac{1}{\xi^2-\frac{Q^2}{m^2_l}\xi+\frac{Q^2}{m^2_l}}\right]
\ln\left[\frac{m^2_l\xi^2}{Q^2(\xi-1)}\right]\,,
\nonumber\\
\nonumber\\
{\mathcal I}_2(Q^2)&=&\int^1_0 {\rm d}\xi\left[\frac{\xi}{\xi^2-\frac{Q^2}{m^2_l}\xi+\frac{Q^2}{m^2_l}}\right]\,,
\nonumber\\
\nonumber\\
{\mathcal I}_3(Q^2)&=&\int^1_0 {\rm d}\xi\left[\frac{\xi}{\xi^2-\frac{Q^2}{m^2_l}\xi+\frac{Q^2}{m^2_l}}\right]^2
\ln\left[\frac{m^2_l\xi^2}{Q^2(\xi-1)}\right]\,,
\nonumber\\
\nonumber\\
{\mathcal I}_4(Q^2)&=&\int^1_0 {\rm d}\xi\left[\frac{\xi}{\left(\xi^2-\frac{Q^2}{m^2_l}\xi+\frac{Q^2}{m^2_l}\right)^2}\right]
\ln\left[\frac{m^2_l\xi^2}{Q^2(\xi-1)}\right]\,,
\nonumber\\
\nonumber\\
{\mathcal I}_5(Q^2)&=&\int^1_0 {\rm d}\xi\left[\frac{1}{\xi^2-\frac{Q^2}{m^2_l}\xi+\frac{Q^2}{m^2_l}}\right]^2
\ln\left[\frac{m^2_l\xi^2}{Q^2(\xi-1)}\right]\,,
\nonumber\\
\nonumber\\
{\mathcal I}_6(Q^2)&=&\int^1_0 {\rm d}\xi\left[\frac{1}{\xi^2-\frac{Q^2}{m^2_l}\xi+\frac{Q^2}{m^2_l}}\right]\, .
\end{eqnarray}
Each of the above integrals are easily obtainable in closed forms using standard techniques or with {\it Mathematica}. 
Since some of these integrals have rather elaborate expressions, we prefer to omit their explicit expressions in 
this communication. Inserting the loop amplitude ${\mathbb I}_{\rm \,seagull}$ into Eq.~\eqref{eq:seagull_amplitude} we 
obtain our final expression for the seagull amplitude,
\begin{eqnarray}
\widetilde{\mathcal M}^{(i)}_{\rm seagull}&=&\frac{\alpha^2}{m^2_l M}\left[{\mathcal N}_1 {\mathcal I}_1 - {\mathcal N}_2 \left({\mathcal I}_2 + \frac{Q^2}{m^2_l}{\mathcal I}_3\right)\right. 
\nonumber\\
&&\left.\, - {\mathcal N}_3\left({\mathcal I}_6 - \frac{Q^2}{m^2_l}{\mathcal I}_5\right)  - {\mathcal N}_4\frac{Q^2}{m^2_l}{\mathcal I}_4\right]\,,
\end{eqnarray}
where ${\mathcal N_i}\not\propto {\mathcal M}_{\gamma}$ ($i=1,...,4$) are defined as
\begin{eqnarray}
{\mathcal N}_1&=&[\bar u(p^\prime)\gamma^\mu(m_l+\slashed{p})\gamma_\mu u(p)][\chi_p^\dagger(p_p^\prime)\chi_p(p_p)]\,,
\nonumber\\
{\mathcal N}_2 &=&[\bar u(p^\prime)\gamma^\mu(\slashed{p}-\slashed{Q})\gamma_\mu u(p)][\chi_p^\dagger(p_p^\prime)\chi_p(p_p)]\,,
\nonumber\\
{\mathcal N}_3 &=&[\bar u(p^\prime)\gamma^\mu \slashed{Q} \gamma_\mu u(p)][\chi_p^\dagger(p_p^\prime)\chi_p(p_p)]\,,
\nonumber\\
{\mathcal N}_4&=&[\bar u(p^\prime)\gamma^\mu(2\slashed{Q}-\slashed{p})\gamma_\mu u(p)][\chi_p^\dagger(p_p^\prime)\chi_p(p_p)] \,.
\end{eqnarray}
Thus, in essence we find that the seagull amplitude, unlike the TPE box amplitudes in SPA, {\it does not} 
naturally factorize into the LO amplitude ${\mathcal M}_{\gamma}$ times a $Q^2$ dependent function $f(Q^2)$. 
This result is consistent with the proposition made in Ref.~\cite{BlundenPRL} that the one-loop virtual 
radiative corrections, and in particular the TPE amplitudes, can be expressed as, 
${\mathcal M}_{\rm 1-Loop}=f(Q^2){\mathcal M}_{\gamma}+\overline{\mathcal M}_{\rm 1-Loop}$. The factorizable 
IR-divergent first term constitutes the dominant, so-called {\it outer corrections}, independent of the hadron 
structure, while the non-factorizable IR-finite second term constitutes small corrections, the so-called 
{\it inner corrections}. In most works these latter corrections are hadron structure dependent and are often 
ignored in ultra-relativistic approximations. In  the HB$\chi$PT approach we find that the dominant TPE box diagrams 
in SPA can be identified with the former corrections, while the seagull term can be identified with the latter 
ones. At the order of our accuracy, the latter corrections are free of low-energy constants and, therefore, 
are hadron structure independent. \\

Finally, it is noteworthy that, while determining $\delta^{\rm (seagull)}_{\gamma\gamma}$, Eq.~\eqref{eq:deltaYY_seagull}, 
the integrals ${\mathcal I}_5$ and ${\mathcal I}_6$ drop out of the calculation due to the vanishing of spin 
trace of ${\mathcal N}_3$ with  ${\cal M}_\gamma$, Eq.~(\ref{eq:Mgamma}). We find 
\begin{eqnarray}
\sum_{spins}{\mathcal M}^*_\gamma {\mathcal N}_3  &=&  0\,, 
\end{eqnarray}

\vspace{-0.2cm}

and

\vspace{-0.2cm}

\begin{eqnarray}
\sum_{spins}{\mathcal M}^*_\gamma\, {\mathcal N}_1 &=& - \sum_{spins}{\mathcal M}^*_\gamma\, {\mathcal N}_2 
= \sum_{spins}{\mathcal M}^*_\gamma\, {\mathcal N}_4 
\nonumber\\
&=&-\frac{16 e^2 m^2_l}{Q^2}(E+E^\prime)(E_p+M)(E_p^\prime+M) 
\nonumber\\
&=& -\frac{128 e^2 m^2_lM^2E}{Q^2}\left[1+{\mathcal O}\left(\frac{1}{M}\right)\right]\, .  
\end{eqnarray}
Here we have used $E^\prime=E+{\mathcal O}(M^{-1})$ and $E^\prime_p=M+{\mathcal O}(M^{-1})$, since 
the seagull diagram already is an NLO amplitude.    



\begin{thebibliography}{10}
\bibitem{Pohl:2010zza}
R.~Pohl, {\it et al.}, 
\newblock Nature \textbf{466} (2010) 213. 

\bibitem{Antognini}
A.~Antognini {\it et al.}, 
\newblock Science {\bf 339} (2013) 417.

\bibitem{Codata}
P.~J.~Mohr {\it et al.} 
\newblock [Codata], Rev. Mod. Phys. {\bf 88} (2016) 035009. 

\bibitem{Pohl:2013yb}
R.~Pohl, {\it et al.},
\newblock Annu. Rev. Nucl. Part. Science, {\bf 63} (2013) 175 . 

\bibitem{Bernauer:2010wm}
J. Bernauer {\it et al.} [A1 Collaboration], 
\newblock Phys. Rev. Lett. {\bf 105} (2010) 242001. 

\bibitem{Zhan:2011ji}
X. Zhan {\it et al.}, 
\newblock Phys. Lett. B {\bf 705} (2011) 59.

\bibitem{Mihovilovic:2016rkr}
M. Mihovilovi\v{c} {\it et al.}, 
\newblock Phys. Lett. B {\bf 771} (2017) 194. 

\bibitem{Bezginov:2019}
N.~Bezginov {\em et~al.},
\newblock Science {\bf 365} (2019) 1007.

\bibitem{Walker:1994}
R.~C.~Walker {\em et~al.},
\newblock Phys. Rev. D {\bf 49} (1994) 5671.

\bibitem{Bosted:1994tm}
P.~E.~Bosted
\newblock Phys. Rev. C {\bf 51} (1995) 409.

\bibitem{Milbrath:1998}
B.~D.~Milbrath {\em et~al.},
\newblock Phys. Rev. Lett.  {\bf 80} (1998) 452.

\bibitem{Milbrath:1999}
B.~D.~Milbrath {\em et~al.},
\newblock Phys. Rev. Lett.  {\bf 82} (1999) 2221(E).

\bibitem{Jones:1999rz}
M.~K.~Jones {\em et~al.},
\newblock Phys. Rev. Lett. {\bf 84} (2000) 1398.

\bibitem{Gayou:2001qt}
O.~Gayou  {\em et~al.},
\newblock Phys. Rev. C {\bf 64} (2001) 038202.

\bibitem{Gayou:2002}
O.~Gayou  {\em et~al.},
\newblock Phys. Rev. Lett. {\bf 88} (2002) 092301.

\bibitem{Brash:2002}
E.~J.~Brash {\em et~al.},
\newblock Phys. Rev. C {\bf 65} (2002) 051001.

\bibitem{Arrington:2003df}
J.~Arrington,
\newblock Phys. Rev. C {\bf 68} (2003) 034325.

\bibitem{Perdrisat:2006hj}
C.~F.~Perdrisat, V.~Punjabi, and M.~Vanderhaeghen, 
\newblock Prog. Part. Nucl. Phys. {\bf 59} (2007) 694.

\bibitem{Arrington:2011dn}
J.~Arrington, P.~G.~Blunden, and W.~Melnitchouk, 
\newblock Prog. Part. Nucl. Phys. {\bf 66} (2011) 782.

\bibitem{Carlson:2015jba}
C.E.~Carlson,
\newblock Prog. Part. Nucl. Phys. {\bf 82} (2015) 59.

\bibitem{Tsai:1961zz}
Y-S.~Tsai, 
\newblock Phys. Rev. {\bf 122} (1961) 1898.

\bibitem{Mo:1968cg}
L.~W.~Mo and Y-S.~Tsai, 
\newblock Rev. Mod. Phys. {\bf 41} (1969) 205.

\bibitem{Maximon:2000hm}
L.~C.~Maximon and J.~A.~Tjon,
\newblock Phys. Rev. C {\bf 62} (2000) 054320.
L.~C.~Maximon, Rev. Mod. Phys. {\bf 41} (1969) 193. 

\bibitem{Vanderhaeghen2000} 
M.~Vanderhaeghen {\it et al.}, 
\newblock Phys. Rev. C {\bf 62} (2000) 025501.

\bibitem{Gramolin2014} 
A.~V.~Gramolin {\it et al.}, 
\newblock J. Phys. G {\bf 41} (2014) 115001. 

\bibitem{Prad} 
A.~Gasparian,  
\newblock [PRad Collaboration], EPJ Web Conf. {\bf 73} (2014) 07006; 
W.~Xiong,  A.~Gasparian {it et al.}, Nature {\bf 575} (2019) 147. 

\bibitem{Prad2}
C.~Peng and H.~Gao,
\newblock  [PRad Collaboration], EPJ Web Conf., {\bf 113} (2016) 03007. 

\bibitem{Gilman:2013eiv}
R.~Gilman {\em et~al.} [MUSE Collaboration] ,
[nucl-ex:1303.2160].

\bibitem{Gilman:2013vma}
R.~Gilman,
\newblock AIP Conf. Proc. {\bf 1563} (2013) 167.

\bibitem{tomalak2014two}
O.~Tomalak and M.~Vanderhaeghen,
\newblock Phys. Rev. D {\bf 90} (2014) 013006.

\bibitem{Afanasev:2017gsk}
A.~Afanasev {\em et~al.},
\newblock Prog. Part. Nucl. Phys. {\bf 95} (2017) 245.

\bibitem{Talukdar:2018hia}
P.~Talukdar, F.~Myhrer, G.~Meher, and U.~Raha,
\newblock Eur. Phys. J. A{\bf 54} (2018) 195.

\bibitem{Myhrer:2018ski}
F.~Myhrer, P.~Talukdar and U.~Raha,
\newblock Few Body Syst. {\bf 59} (2018) 62,

\bibitem{McKinley:1948zz}
W.~A.~McKinley and H.~Feshbach, 
\newblock Phys. Rev. {\bf 74} (1948) 1759.

\bibitem{Carlson:2007sp}
C.~E.~Carlson and M.~Vanderhaeghen, 
\newblock Ann. Rev. Nucl. Part. Sci. {\bf 57} (2007) 171.

\bibitem{Borisyuk:2006fh}
D.~Borisyuk and A.~Kobushkin, 
\newblock Phys. Rev. C {\bf 74} (2006) 065203.

\bibitem{Borisyuk:2008es}
D.~Borisyuk and A.~Kobushkin, 
\newblock Phys. Rev. C {\bf 78} (2008) 025208.

\bibitem{Borisyuk:2010cv}
D.~Borisyuk and A.~Kobushkin, 
\newblock Phys. Rev. C {\bf 83} (2011) 025203.

\bibitem{Gorchtein:2006mq}
M.~Gorchtein,
\newblock Phys. Lett. B {\bf 644} (2007) 322.

\bibitem{Kuraev:2008gw}
E.~A.~Kuraev and E.~Tomasi-Gustafsson,
\newblock Phys. Part. Nucl. Lett. {\bf 7} (2010) 67.

\bibitem{Belushkin:2006qa}
M.~A.~Belushkin, H.~W.~Hammer, and Ulf.-G.~Mei{\ss}ner,
\newblock Phys. Rev. C {\bf 75} (2007) 035202.

\bibitem{Belushkin:2007zv}
M.~A.~Belushkin, H.~W.~Hammer, and Ulf.-G.~Mei{\ss}ner,
\newblock Phys. Lett. B {\bf 658} (2008) 138.

\bibitem{Lorenz:2012tm}
I.~T.~Lorenz, H.~W.~Hammer, and Ulf.-G.~Mei{\ss}ner
\newblock Eur. Phys. J. A {\bf 48} (2012) 151.

\bibitem{Hoferichter:2016duk}
M.~Hoferichter {\em et~al.},
\newblock Eur. Phys. J. A {\bf 52} (2016) 331.

\bibitem{tomalak2015three}
O.~Tomalak and M.~Vanderhaeghen, 
\newblock Eur. Phys. J. A {\bf 51} (2015) 24.

\bibitem{Tomalak:2015aoa}
O.~Tomalak and M.~Vanderhaeghen,
\newblock Phys. Rev. D {\bf 93} (2016) 013023.

\bibitem{Tomalak:2015hva}
O.~Tomalak and M.~Vanderhaeghen,
\newblock Eur. Phys. J. C {\bf 76} (2016) 125.   

\bibitem{tomalak2018}
O.~Tomalak and M.~Vanderhaeghen,
\newblock Eur. Phys. J. C {\bf 78} (2018) 514; 

\bibitem{Tomalak:2017npu}
O.~Tomalak, 
\newblock Eur. Phys. J. C {\bf 77} (2017) 858.

\bibitem{Kondratyuk:2001qu}
S.~Kondratyuk and O.~Scholten,
\newblock Phys. Rev. C {\bf 64} (2001) 024005.

\bibitem{Penner:2002md}
G.~Penner and U.~ Mosel,
\newblock Phys. Rev. C {\bf 66} (2002) 055212.

\bibitem{Korchin:2003ah}
A.~{\relax Yu}.~Korchin and  and O.~Scholten, 
\newblock Phys. Rev. C {\bf 68} (2003) 045206.

\bibitem{Kondratyuk:2005kk}
S.~Kondratyuk  {\em et~al.},
\newblock Phys. Rev. Lett. {\bf 95} (2005) 172503. 

\bibitem{Kondratyuk:2007hc}
S.~Kondratyuk and P.~G.~Blunden,
\newblock Phys. Rev. C {\bf 75} (2007) 038201.

\bibitem{Zhou:2009nf}
H.~Q.~Zhou {\em et~al.},
\newblock Phys. Rev. C {\bf 81} (2010) 035208.

\bibitem{BlundenPRL}
P.~G.~Blunden, W.~Melnitchouk and J.~A.~Tjon, 
\newblock Phys. Rev. Lett. {\bf 91} (2003) 142304.

\bibitem{Koshchii:2017dzr}
O.~Koshchii and A.~Afanasev,
\newblock Phys. Rev. D {\bf 96} (2017) 016005.

\bibitem{Bernard:1995dp}
V.~Bernard, N.~Kaiser, and U.-G. Mei{\ss}ner,
\newblock Int. J. Mod. Phys. E{\bf 4} (1995) 193.

\bibitem{Bernard:2007}
V. Bernard, 
\newblock Prog. Part. Nucl. Phys. {\bf 60} (2008) 82. 

\bibitem{Talukdar}
P. Talukdar {\it et al.} in preparation.

\bibitem{Sick:1998cvq}
I.~Sick and D.~Trautmann,
\newblock Nucl. Phys A {\bf 637} (1998) 559.

\bibitem{Blunden:2005jv}
P.~G.~Blunden and I.~ Sick,
\newblock Phys. Rev. C {\bf 72} (2005) 057601.

\bibitem{hooft:1979}
G.~'t~Hooft and M.~Veltman,
\newblock Nucl. Phys B {\bf 153} (1979) 365.

\bibitem{passarino:1979}
G.~Passarino and M.~Veltman,
\newblock Nucl. Phys B {\bf 160} (1979) 151.

\bibitem{beneke:1998}
M.~Beneke and V.~A.~Smirnov,
\newblock Nucl. Phys B {\bf 522} (1998) 321.

\bibitem{grozin:2001}
A.~I.~Davydychev and A.~G.~Grozin,
\newblock Eur. Phys. J. C {\bf 20} (2001) 333.

\bibitem{zupan:2002}
J.~Zupan,
\newblock Eur. Phys. J. C {\bf 25} (2002) 233.

\end{thebibliography}
\end{document}